\documentclass[prstab,11pt,superscriptaddress,showpacs,aps]{revtex4-1}

\usepackage{graphicx}
\usepackage{booktabs}
\usepackage{amssymb}
\usepackage{amsmath}
\usepackage{dcolumn}
\usepackage{bm}% bold math
\usepackage{hyperref}% add hypertext capabilities

\newcommand{\bild}[4]{\begin{figure}\begin{center}\includegraphics[width=#1mm]{#2}\caption[]{\label{#3}#4}\end{center}\end{figure}}

\newif\ifpdf
\ifx\pdfoutput\undefined
   \pdffalse
\else
  \pdfoutput=1
   \pdftrue
\fi
\ifpdf
   \usepackage{graphicx}
   \usepackage{epstopdf}
\else
   \usepackage{graphicx}
\fi

\makeatletter
\def\tagform@#1{\maketag@@@{\ignorespaces#1\unskip\@@italiccorr}}
\let\orgautoref\autoref
\providecommand{\Autoref}[1]
{
\def\equationautorefname{Equation}\def\figureautorefname{Figure}\def\sectionautorefname{Section}\def\tableautorefname{Table}\def\subfigureautorefname{Figure}\orgautoref{#1}
}
\renewcommand{\autoref}[1]
{
\def\equationautorefname{Eq.}\def\figureautorefname{Fig.}\def\sectionautorefname{Sec.}\def\tableautorefname{Tab.}\def\subfigureautorefname{Fig.}\orgautoref{#1}
}

\begin{document}

%\preprint{APS/123-QED}

\title{Interpretation of transverse tune spectra in a heavy-ion synchrotron at high intensities}
\thanks{This work is supported by DITANET (novel DIagnostic Techniques for future particle Accelerators: A Marie Curie Initial Training NETwork), Project Number ITN-2008-215080}

\author{R. Singh}
\affiliation{GSI Helmholzzentrum f\"ur Schwerionenforschung mbH, Planckstrasse 1, 64291 Darmstadt, Germany}
\affiliation{Technische Universit\"at Darmstadt, Schlossgartenstr.8, 64289 Darmstadt, Germany}
\author{O. Boine-Frankenheim}
\affiliation{GSI Helmholzzentrum f\"ur Schwerionenforschung mbH, Planckstrasse 1, 64291 Darmstadt, Germany}
\affiliation{Technische Universit\"at Darmstadt, Schlossgartenstr.8, 64289 Darmstadt, Germany}
\author{O. Chorniy}
\author{P. Forck}
\author{R. Haseitl}
\author{W. Kaufmann}
\author{P. Kowina}
\author{K. Lang}
\affiliation{GSI Helmholzzentrum f\"ur Schwerionenforschung mbH, Planckstrasse 1, 64291 Darmstadt, Germany}
\author{T. Weiland}
\affiliation{Technische Universit\"at Darmstadt, Schlossgartenstr.8, 64289 Darmstadt, Germany}

\date{\today}% It is always \today, today,

\begin{abstract}
Two different tune measurement systems have been installed in the GSI heavy-ion synchrotron SIS-18. Tune spectra are obtained with high accuracy using these fast and sensitive systems . Besides the machine tune,
the spectra contain information about the intensity dependent coherent tune shift 
and the incoherent space charge tune shift. The space charge tune shift is derived from a fit of the 
observed shifted positions of the synchrotron satellites to an analytic expression for the head-tail eigenmodes with space charge. Furthermore, the chromaticity is extracted from the measured head-tail mode structure.      
The results of the measurements provide experimental evidence of the importance of 
space charge effects and head-tail modes for the interpretation of transverse beam signals at high intensity. 
\end{abstract}

\maketitle

\section{Introduction}\label{sec:level1}
Accurate measurements of the machine tune and of the chromaticity are of importance for the operation of fast ramping, high intensity ion synchrotrons.
In such machines the tune spread $\delta Q_{x,y}$ at injection energy due to space charge and chromaticity 
can reach values as large as 0.5. In order to limit the incoherent particle tunes to the resonance free region the machine tune has be controlled with a precision better than $\Delta Q\approx 10^{-3}$.
In the GSI heavy-ion synchrotron SIS-18 there are currently two betatron tune measurement systems installed.
The frequency resolution requirements of the systems during acceleration are specified as $10^{-3}$, but they provide much higher resolution ($10^{-4}$) on injection and extraction plateaus.
The Tune, Orbit and Position measurement system
(TOPOS) is primarily a digital position measurement system which calculates the tune from the measured position~\cite{Kowina2010}. The Baseband Q measurement system (BBQ) conceived at CERN performs a tune measurement based on the concept of diode based bunch envelope detection~\cite{Gasior2012}. The BBQ system provides a 
higher measurement sensitivity than the TOPOS system.
Passive tune measurements require high sensitivity (Schottky) pick-ups, low noise electronics and long averaging time to achieve reasonable signal-to-noise ratio. For fast tune measurements with the TOPOS and BBQ systems, which both use standard pick-ups, the beam has to be excited externally in order to measure the transverse beam signals. 

For low intensities the theory of transverse signals from bunched beams and the tune measurement principles 
from Schottky or externally excited signals 
are well known~\cite{Chattopadhyay1984,Boussard1995}. In intense, low energy bunches the transverse signals 
and the tune spectra can be modified significantly by the transverse space charge force and by ring impedances.
Previously, the effect of space charge on head-tail modes had been the subject of several analytical and simulation 
studies \cite{Blaskiewicz1998,Boine-Frankenheim2009,Burov2009,Balbekov2009,Kornilov2010}. 

Recently, the modification of transverse signals from high intensity bunches was observed in the SIS-18
for periodically excited beams \cite{Singh2012} and for initially kicked bunches \cite{Kornilov2012},
where the modified spectra was explained in terms of the space charge induced head-tail mode shifts. 

This contribution aims to complement the previous studies and 
extract the relevant intensity parameters from tune spectra measurements using the TOPOS and BBQ tune measurement systems.~\Autoref{sec:bunched-btf} presents the frequency content of transverse bunched beam signals briefly.~\Autoref{sec:ht-modes} presents the space charge and image current effects on tune measurements and respective theoretical models.~\Autoref{sec:meas-setup} report on the experimental conditions and compares the characteristics of two installations as well as the various excitation methods.~\Autoref{sec:ex-obser} presents the experimental results in comparison with the theoretical estimates of various high intensity effects.

\section{Transverse bunch signals}\label{sec:bunched-btf}

Theoretical and experimental work related to transverse Schottky signals and beam transfer functions (BTFs) for bunched beam 
at low intensities can be found in the existing literature ~\cite{Chattopadhyay1984,Sacherer1968,Linnecar1981}. 
The transverse signal of a beam is generated by the beam's dipole moment
\begin{equation}
d(t)=\sum_{j=1}^{N}x_j(t)I_j(t)
\end{equation}
where $x_j(t)$ is the horizontal/vertical offset and $I_j(t)$ the current of the $j^{th}$ 
particle at the position of a pick-up (PU) in the ring. 
The sum extends over all $N$ particles in the detector.
The Schottky noise power spectrum as a function of the frequency $f$ is defined as $S(f)=| d(f) |^2$
where $d(f)$ is the Fourier transformed PU signal.
For a beam excited by an external amplitude spectrum $G(f)$, the transverse response function is defined as $r(f)=d(f)/G(f)$ \cite{Borer1979}. 

If the transverse signal from a low intensity bunch is sampled with the revolution period $T_s$, then the positive frequency spectrum 
consists of one set of equidistant lines 
\begin{equation}
Q_{k}=Q_0+\Delta Q_{k}, 
\end{equation}
usually defined as baseband tune spectrum,
where $Q_0$ is the fractional part of the machine tune, $\Delta Q_{k}=\pm kQ_s$ are the synchrotron satellites  
and $Q_s$ is the synchrotron tune.  

For a single particle performing betatron and synchrotron oscillations, the relative amplitudes of the 
satellites are \cite{Chattopadhyay1984}
\begin{equation}\label{eq:rel-amp}
\mid d_k\mid\sim \mid J_k(\chi/2)\mid
\end{equation}
where $\chi=2\xi\phi_m/\eta_0$ is the chromatic phase, $\xi$ is the chromaticity, $\phi_m$ is the longitudinal oscillation amplitude of the particle
and $\eta_0$ the frequency slip factor. $J_k$ are the Bessel functions of order $k$. 
In bunched beams, the relative height and width of the lines depends on the bunch distribution.
The relative height is also affected by the characteristics of the external noise excitation 
or by the initial transverse perturbation applied to the bunch.
In the absence of transverse nonlinear field components 
the width of each satellite with $k\ne 0$ is determined   
by the synchrotron tune spread $\delta Q_k\approx |k|Q_s \phi^2_m/16$. 
An example tune spectrum obtained from a simulation code for a Gaussian bunch distribution
is shown in~\autoref{fig:patric_single_q0}.

\bild{85}{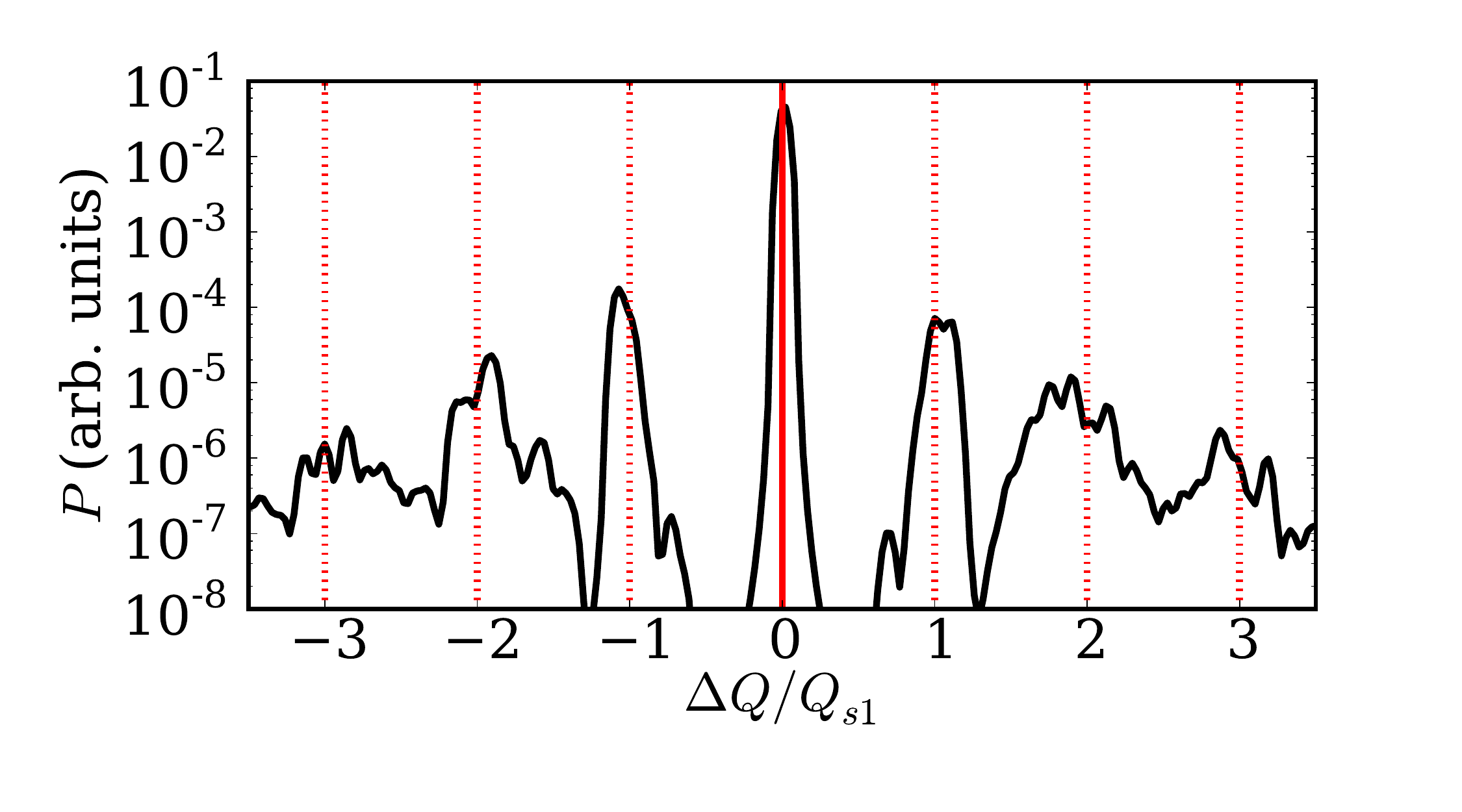}{fig:patric_single_q0}
{Baseband tune spectrum from a simulation code for a low intensity bunch.}

\section{Tune spectrum for high intensity}\label{sec:ht-modes}

At high beam intensities, the transverse space charge force together with the coherent 
force caused by the beam pipe impedance will affect the motion of the beam particles and also the tune spectrum. 
The space charge force induces an incoherent tune shift $Q_0-\Delta Q_{sc}$ for a symmetric beam profile of homogeneous density where
\begin{equation}
\label{eq:sctune}  \Delta Q_{sc}=\frac{qI_p R}{4\pi\epsilon_0 c E_0 {\gamma_0}^{2}\beta_0^3\varepsilon_x}
\end{equation}
is the tune shift, $I_p$ the bunch peak current, $q$ the particle charge and $E_0=\gamma_0 mc^2$ the total energy. 
The relativistic parameters are $\gamma_0$ and $\beta_0$, the ring radius is $R$ and the emittance of the rms equivalent K-V distribution is $\varepsilon_x$.
%For a rms-equivalent transverse Gaussian profile, with a rms emittance $\tilde{\varepsilon}_x=\varepsilon_x/4$,the maximum space-charge tune shift is larger by a factor 2. This can be expressed in terms of a form factor $g_f=2$ in Eq. \ref{eq:sctune}. 
In the case of an elliptic transverse cross-section the emittance $\varepsilon_x$ in~\autoref{eq:sctune} should be replaced by
\begin{equation} 
 \frac{1}{2}\left( \varepsilon_x + \sqrt{\varepsilon_x\varepsilon_y  \frac{Q_{0x}}{Q_{0y}} } \right)
\end{equation}
For the vertical plane the procedure is the same, with $x$ replaced by $y$. 
The image currents and image charges induced in the beam pipe, assumed here to be perfectly conducting, cause a purely imaginary horizontal impedance
 \begin{equation}\label{eq:impedance}
 Z_{x}=-i\frac{Z_0}{2\pi(\beta_0\gamma_0 b_x)^2}
\end{equation}
and real coherent tune shift
\begin{equation}\label{eq:coh_shift}
\Delta Q_{c}=-i\frac{qIR^2 Z_{x}}{2Q_{x0}\beta_0 E_0}
\end{equation}
For a round beam profile with radius $a$ and pipe radius $b$ the coherent tune shift is smaller by 
$\Delta Q_c=\frac{a^2}{b^2}\Delta Q_{sc}$ than the space charge tune shift. Therefore 
the contribution of the pipe is especially important for thick beams $(a\sim b)$ at low or medium energies.

In the presence of incoherent space charge, represented by the tune shift $\Delta Q_{sc}$,
or pipe effects, represented by the real coherent tune shift $\Delta Q_{c}$, 
the shift of the synchrotron satellites in bunches can be reproduced rather well by \cite{Boine-Frankenheim2009}
\begin{equation}\label{eq:htmodes}
\Delta Q_{k}=- \frac{{\Delta Q_{sc}+\Delta Q_{c}}}{2} \pm \sqrt {(\Delta Q_{sc}-\Delta Q_{c})^2/4 + (kQ_s )^2 }
\end{equation}
where the sign $+$ is used for $k>0$. For k=0 one obtains $\Delta Q_{k=0}=-\Delta Q_{c}$. 
The above expression represents the head-tail eigenmodes for an airbag bunch distribution in a barrier potential 
\cite{Blaskiewicz1998} with the eigenfunctions
\begin{equation}\label{eq:eig_func}
\bar{x}(\phi)=\cos(k\pi\phi/\phi_b)\exp(-i\chi\phi/\phi_b)
\end{equation}
where $\bar{x}$ is the local transverse bunch offset, $\chi=\xi\phi_b/\eta_0$ is the chromatic phase, $\phi_b$ is
the full bunch length and $\eta_0$ is slip factor. 
The head-tail mode frequencies obtained from~\autoref{eq:htmodes} are 
shown in~\autoref{fig:ht-modes}.
In Ref.~\cite{Blaskiewicz1998} the analytic solution for the eigenvalues 
\autoref{eq:htmodes} is obtained from a simplified approach, 
where the transverse space charge force is assumed to be constant for all particles.
This assumption is correct if there are only dipolar oscillations. 
In Ref.~\cite{Boine-Frankenheim2009} it is has been pointed out, 
that in the presence of space charge there is an additional envelope 
oscillation amplitude. For the negative-$k$ eigenmodes the envelope contribution
dominates and therefore those modes disappear from the tune spectrum.
In Ref.~\cite{Boine-Frankenheim2009}~\autoref{eq:htmodes} has been successfully compared to Schottky spectra obtained from 3D self-consistent simulations for realistic bunch distributions in rf buckets. 
Analytic and numerical solutions for Gaussian and other bunch distribution valid for $q_{sc}\gg 1$ were presented in~\cite{Burov2009,Balbekov2009}. 

In an rf bucket the synchrotron tune $Q_s$ is a function of the synchrotron oscillation amplitude~$\hat{\phi}$.
For short bunches $Q_s$ corresponds to the small-amplitude synchrotron tune
\begin{equation}
Q_{s0}^2 =\frac{q V_0 h |\eta_0|}{2 \pi m \gamma \beta^2 c^2}
\end{equation}
where $V_0$ is the rf voltage amplitude and $h$ is the rf harmonic number. 

\bild{85}{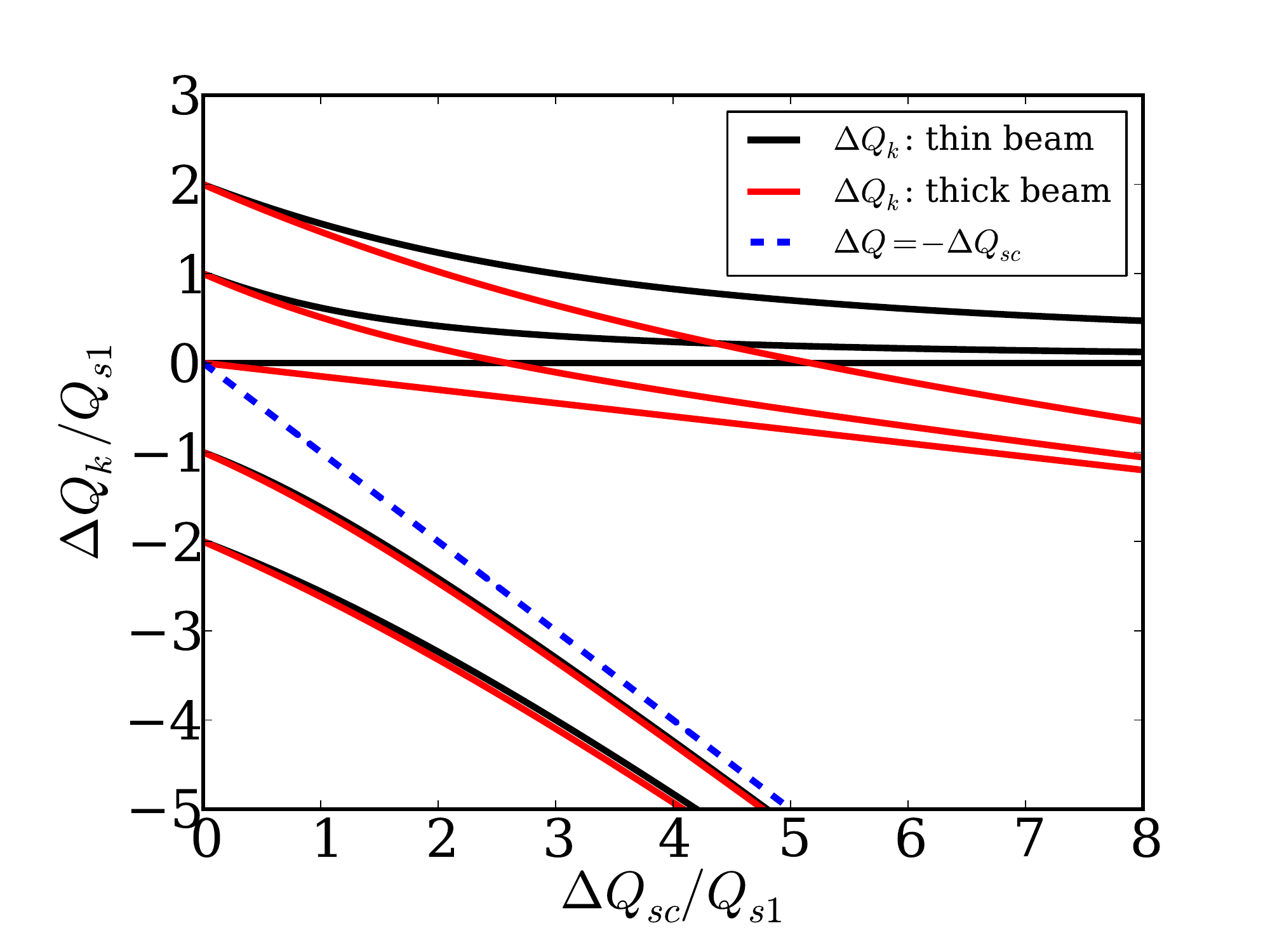}{fig:ht-modes}{Head-tail mode frequencies as a function of the space charge parameter using~\autoref{eq:htmodes}. The red curves represent the result obtained for $q_c=q_{sc}/10$.}

For head-tail modes the space charge parameter is defined as a ratio of the space-charge tune shift 
(~\autoref{eq:sctune}) to the small-amplitude synchrotron tune,
\begin{equation}
\label{eq:scpara}q_{sc} = \frac{\Delta Q_{sc}}{Q_{s0}}
\end{equation}
and the coherent intensity parameter as,
\begin{equation}
\label{eq:cpara}q_{c} = \frac{\Delta Q_{c}}{Q_{s0}}
\end{equation}

An important parameter for head-tail bunch oscillations in long bunches is the effective synchrotron frequency 
which will be different from the small-amplitude synchrotron frequency in short bunches. 
For an elliptic bunch distribution (parabolic bunch profile) with the bunch half-length $\phi_m = \sqrt{5} \sigma_l$ (rms bunch length $\sigma_l$), 
one obtains the approximate analytic expression for the longitudinal dipole tune \cite{boine_rf2005},
\begin{equation}
\frac{Q_{s1}}{Q_{s0}} =\sqrt{1 - \frac{\sigma_l^2}{2}}
\end{equation}
Using $Q_{s1}$ instead of $Q_{s0}$ in Eq. \ref{eq:htmodes} shows a much better agreement 
with the simulation spectra for long bunches in rf buckets (see Ref. \cite{Kornilov2012}). 

\bild{85}{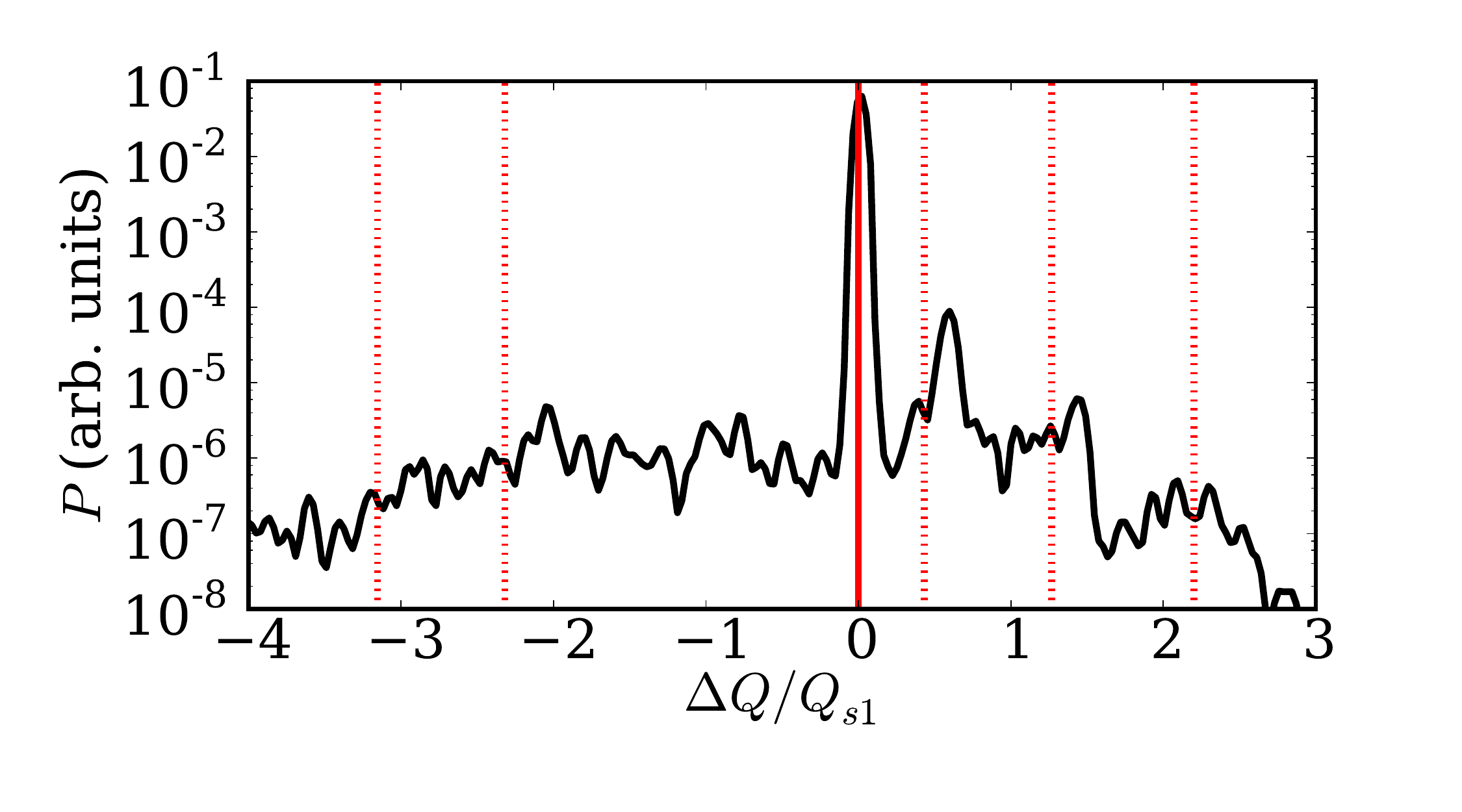}{fig:patric_single_q2}{Simulation result for $q_{sc}=2$.}
\bild{85}{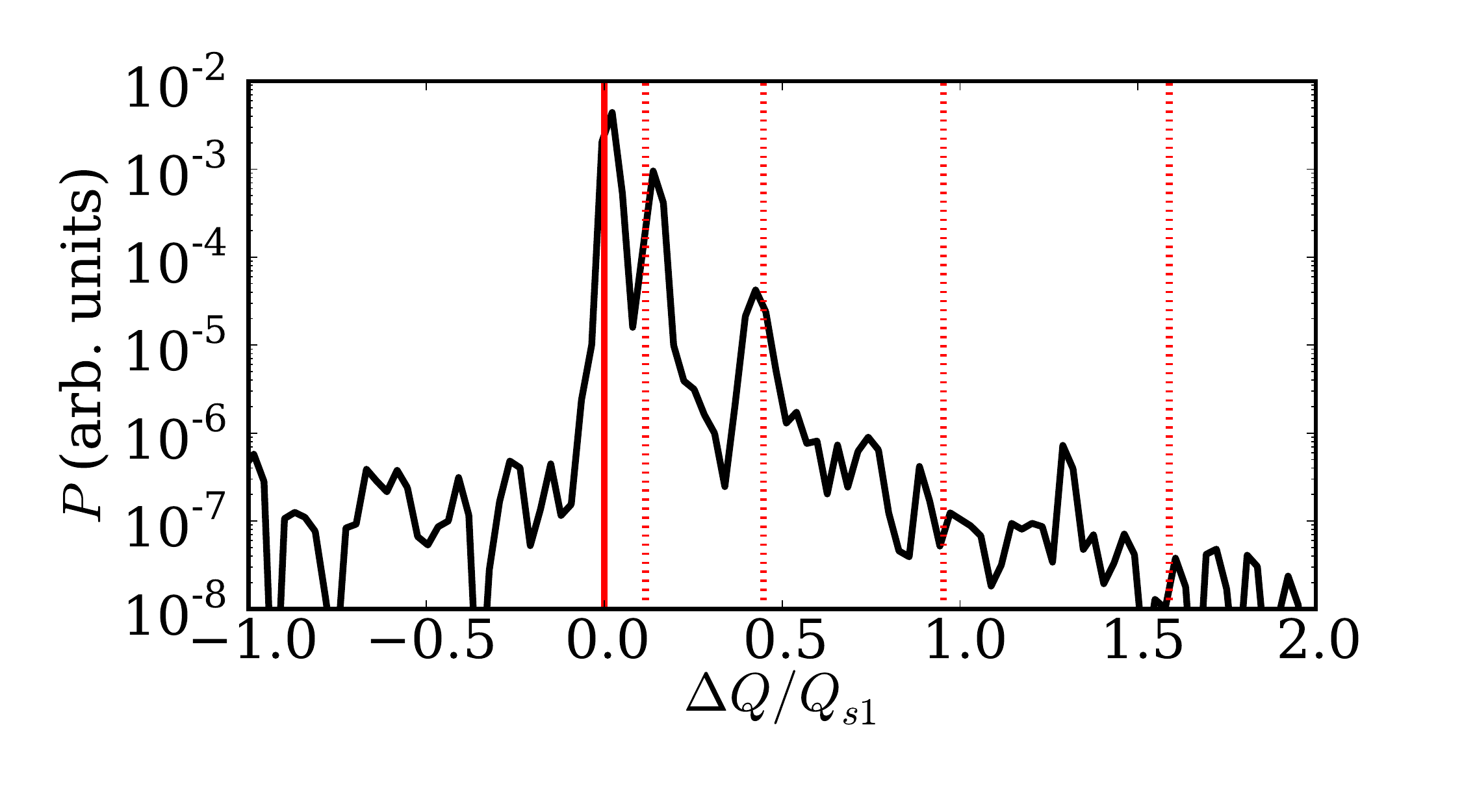}{fig:patric_single_q9}{Simulation result for $q_{sc}=10$.}

For Gaussian bunches with a bunching factor $B_f=0.3$ ($B_f=I_0/I_p$, $I_0$ is the dc current), the transverse tune spectra obtained from PATRIC simulations \cite{Boine-Frankenheim2009} for different space charge factors 
and thin beams ($q_c=0$) are shown in Figures~\ref{fig:patric_single_q0}, \ref{fig:patric_single_q2} and \ref{fig:patric_single_q9}. 
The dotted vertical lines indicate the positions of the head-tail tune shifts obtained from~\autoref{eq:htmodes}     
with $Q_s=Q_{s1}$. For the low-$k$ satellites there is a good agreement between~\autoref{eq:htmodes} and the simulation results. Lines with $k>2$ can only barely be identified in the simulation spectra. 
The positions of the satellites for $k=0,1,2$ together with the predicted head-tail tune shifts from~\autoref{eq:htmodes} are shown in~\autoref{fig:patric_thin}. The error bars indicate 
the obtained widths of the peaks in the tune spectra.

\bild{85}{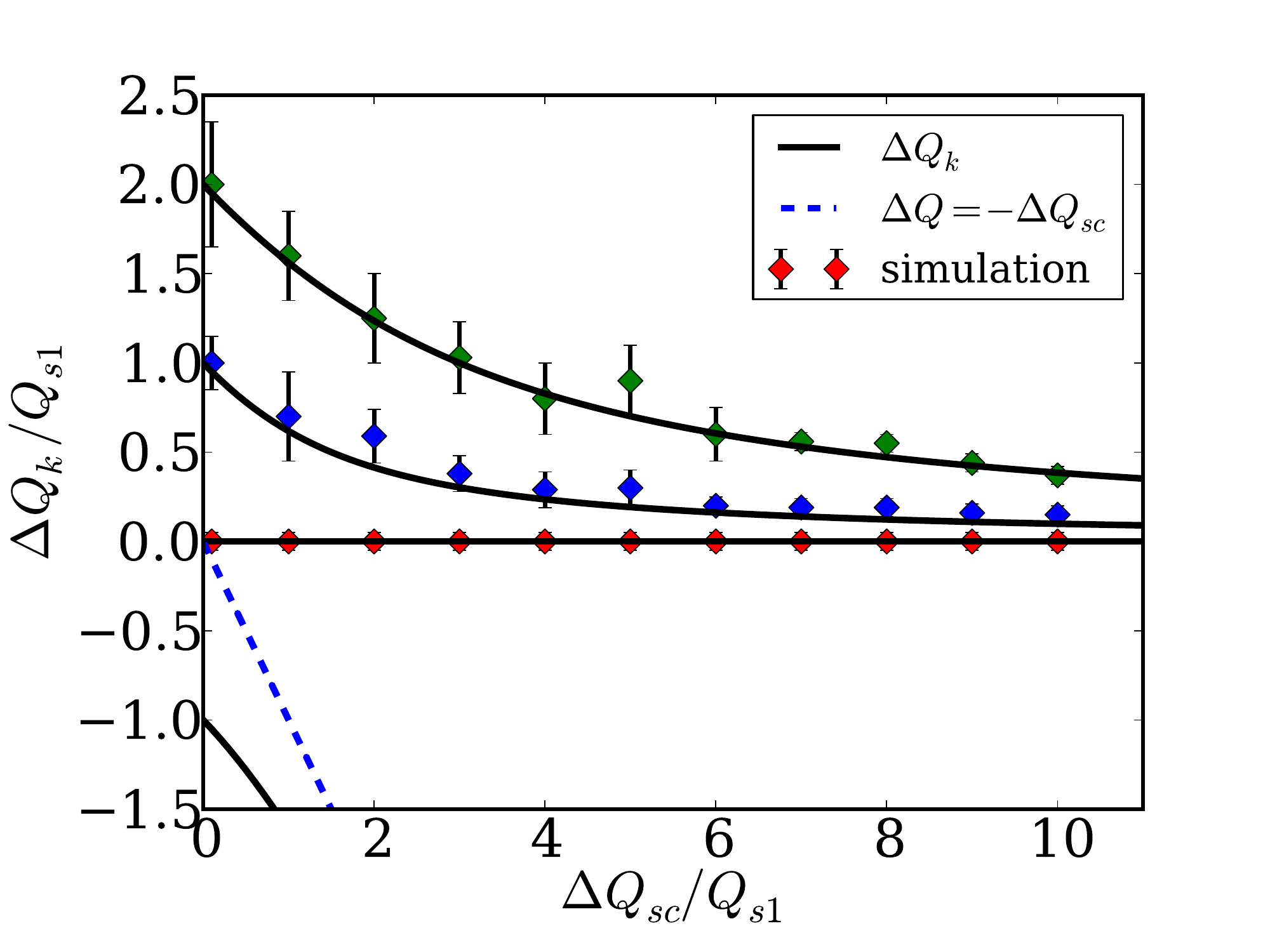}{fig:patric_thin}{Head-tail tune shifts and their width obtained from the simulations for $q_{c}=0$. The error bars indicate the widths of the lines.}

It is important to notice that the simulations for moderate space charge parameters ($q_{sc}\lesssim 10$)
require a 2.5D self-consistent space charge solver. 
The theoretical studies rely on the solution of the M\"ohl-Sch�nauer equation \cite{Moehl1995}, which
assumes a constant space charge tune shift for all transverse particle amplitudes. In contrast to the
self-consistent results, PATRIC simulation studies using the M\"ohl-Sch�nauer equation gave 
tune spectra with pronounced, thin satellites also for large $k$.
In order to account for the intrinsic damping of head-tail modes \cite{Burov2009,Balbekov2009, Kornilov2010}, which
is the main cause of the peak widths obtained from the simulations, a self-consistent treatment is required. 

For thick beams (here $q_c=0.15q_{sc}$, which corresponds to the conditions at injection in the SIS-18)
the positions of the synchrotron satellites obtained from the simulation are indicated in
~\autoref{fig:patric_thick}. Again, the error bars indicate the widths of the peaks. From the plot 
we notice an increase in the spacing between the $k=0, 1, 2$ satellites, relative
to the analytic expression. Also the peak width for $k=1,2$ does not shrink with increasing $q_{sc}$.   

\bild{85}{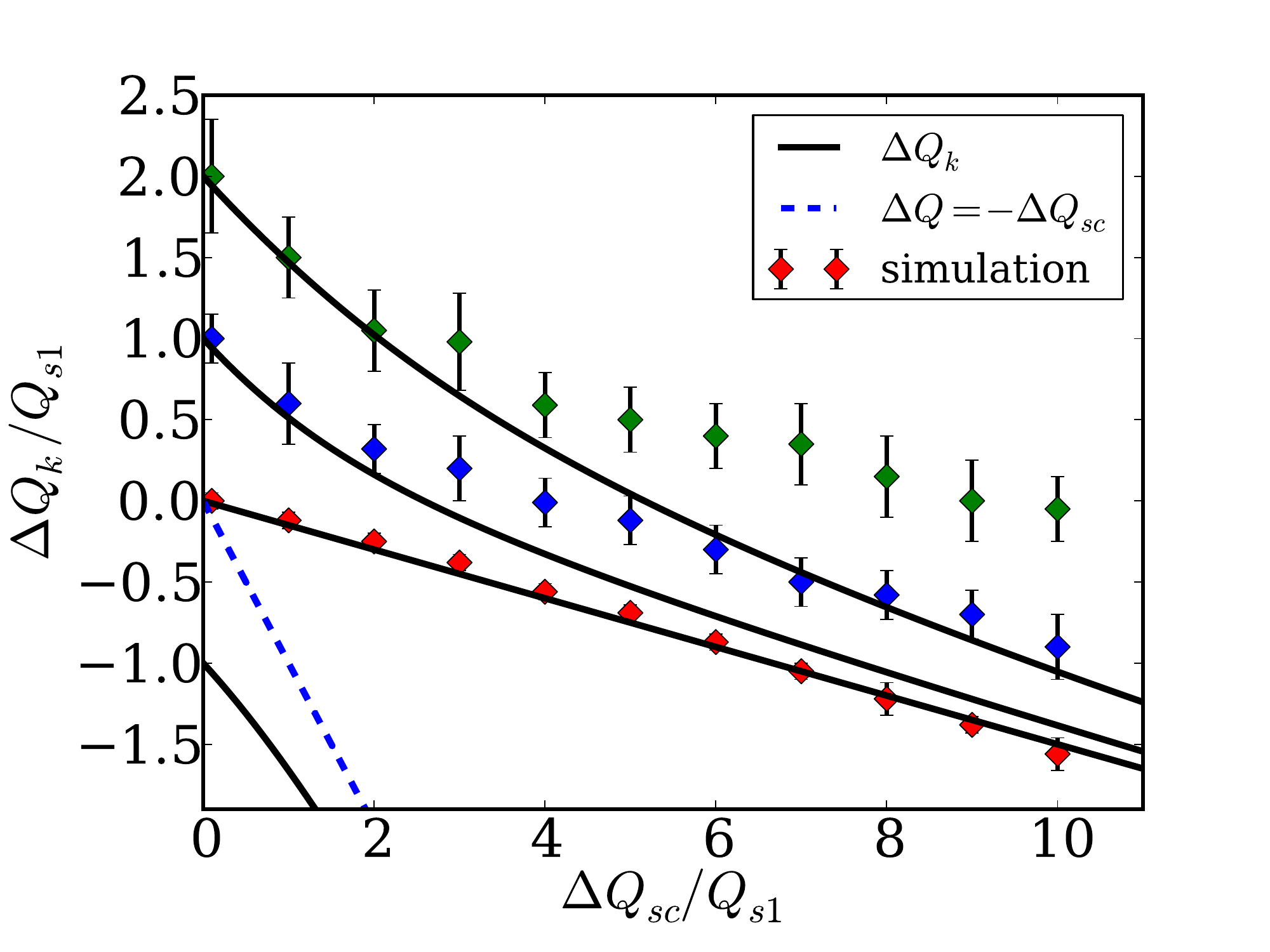}{fig:patric_thick}{Head-tail tune shifts and their width obtained from the simulations for $q_{c}=0.15q_{sc}$. The error bars indicate the widths of the peaks.}

The tune spectra obtained for $q_{sc}=3$, $q_{sc}=5$ and $q_{sc}=10$ 
are shown in~\autoref{fig:patric_single_q3qc015},~\autoref{fig:patric_single_q5qc015} and~\autoref{fig:patric_single_q10qc015}. 
One can observe that for thick beams (here $a\approx 0.4b$) the $k=1$ peak remains very broad up to $q_{sc}=10$.

\bild{85}{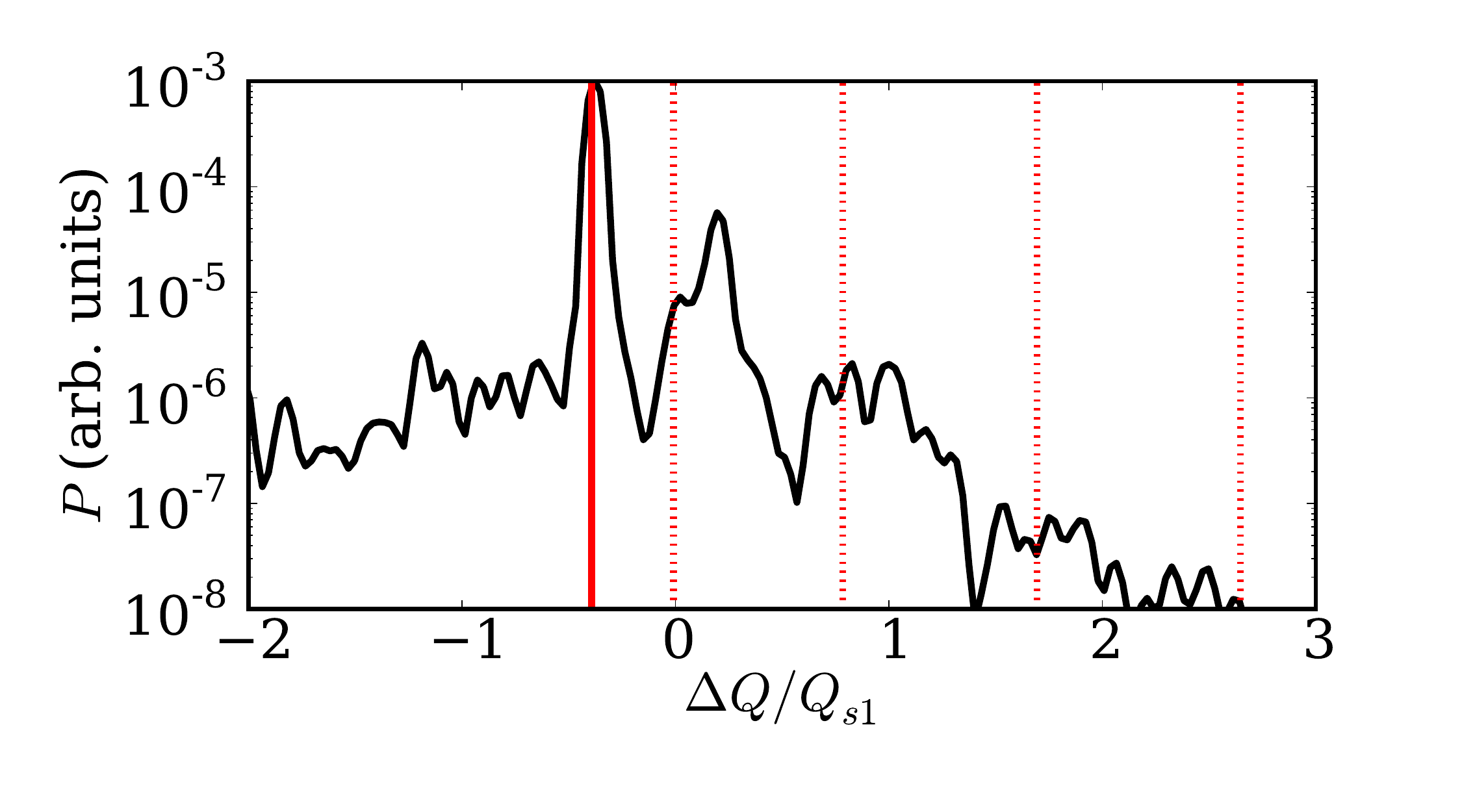}{fig:patric_single_q3qc015}{Tune spectrum obtained from the simulation for $q_{sc}=3$ and $q_c=0.15 q_{sc}$.}

\bild{85}{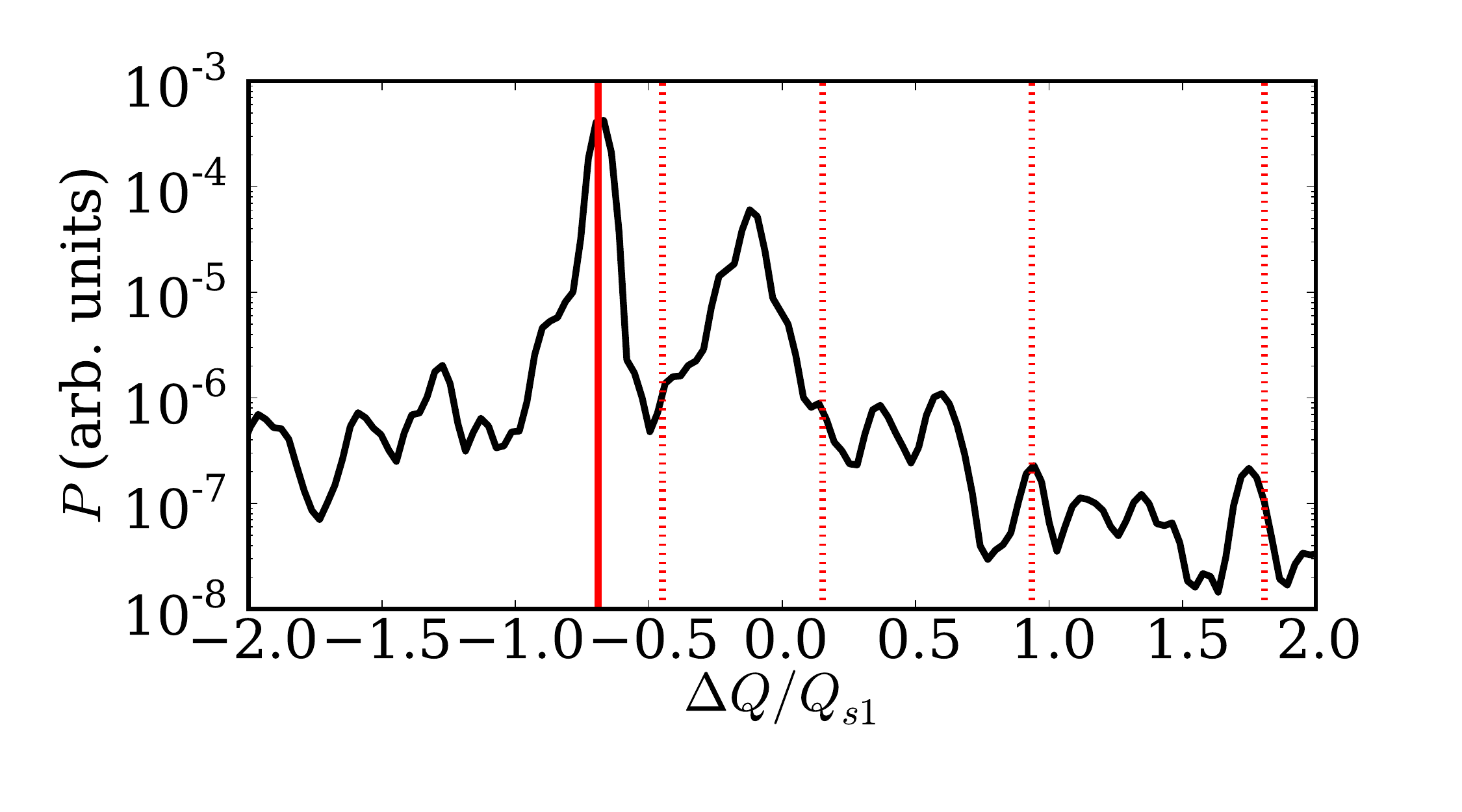}{fig:patric_single_q5qc015}{Tune spectrum obtained from the simulation for $q_{sc}=5$ and $q_c=0.15 q_{sc}$.}

\bild{85}{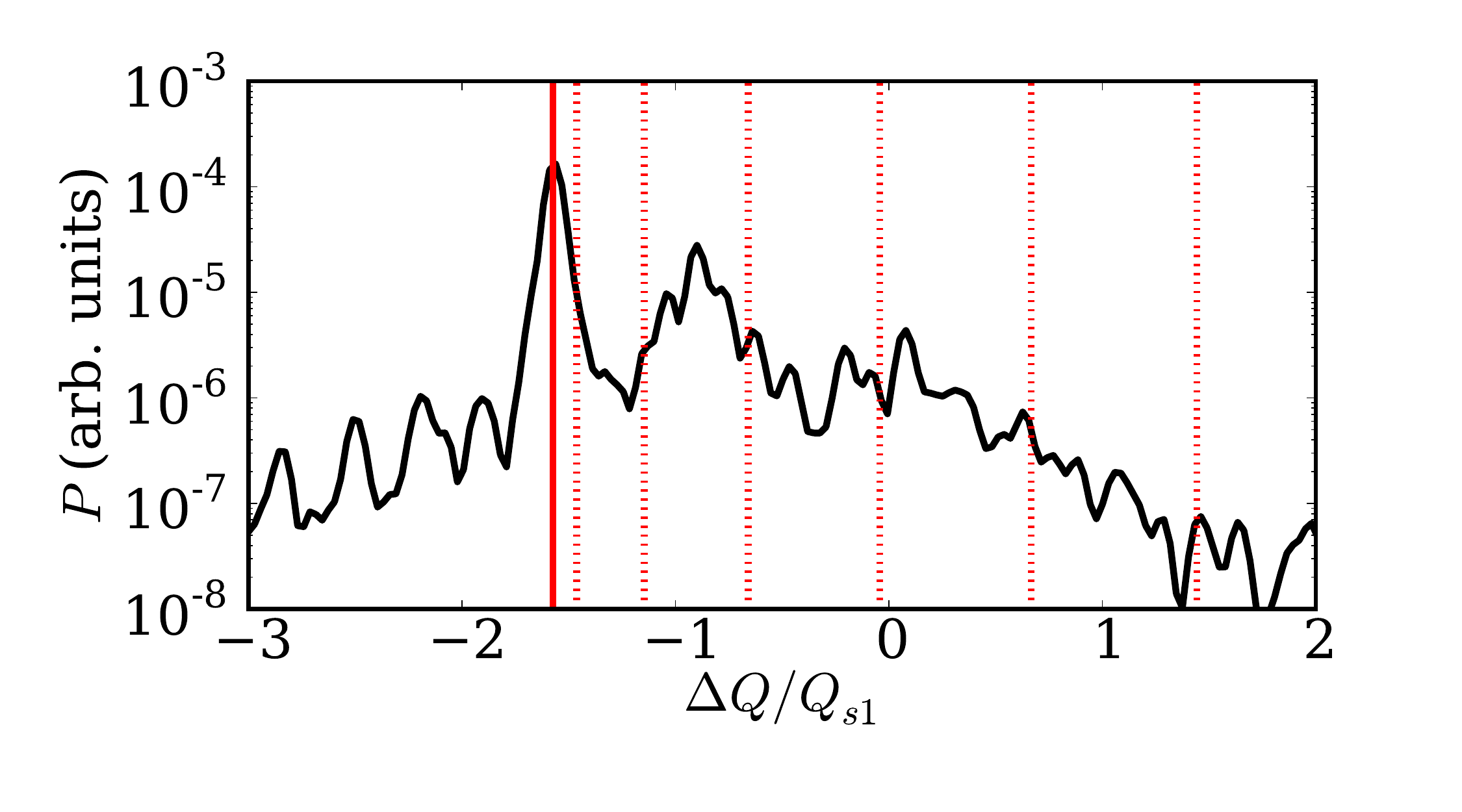}{fig:patric_single_q10qc015}{Tune spectrum obtained from the simulation for $q_{sc}=10$ and $q_c=0.15 q_{sc}$.}

This observation is consistent with a very simplified picture for the upper $q_{sc}$ threshold
for the intrinsic Landau damping of head-tail modes. 
For a Gaussian bunch profile, the maximum incoherent tune shift, including the modulation due to the synchrotron oscillation is 
\begin{equation}
\Delta Q_{\max}=-\Delta Q_{sc}+kQ_s 
\end{equation}
where $\Delta Q_{sc}$ is determined by Eq. \ref{eq:sctune}.
The minimum space charge tune shift is (see Refs. \cite{Balbekov2009,Kornilov2010})
\begin{equation}
\Delta Q_{\min}=-\alpha q_{sc} Q_s+kQ_s 
\end{equation}
where $\alpha$ is determined from the average of the space charge tune shift along a synchrotron oscillation
with the amplitude $\hat{\phi}=\phi_m$. For a parabolic bunch we obtain $\alpha=0.5$. For a Gaussian bunch and $\hat{\phi}=3\sigma_l$, we obtain $\alpha=0.287$. Each band of the incoherent transverse spectrum has a lower boundary determined by the maximum tune shift $\Delta Q_{\max}$ and an upper boundary determined by $\Delta Q_{\min}$. Landau damping, in its very approximate treatment, requires an overlap of the coherent peak with the incoherent band. The head-tail tune for low $q_{sc}$ can be approximated as
\begin{equation}\Delta Q_k=-\frac{1}{2}\left(\Delta Q_{sc}+\Delta Q_c\right)+kQ_s\end{equation}
The distance between the coherent peak and the upper boundary of the incoherent band for fixed $k$ is  \begin{equation}\delta Q_k=\left(\frac{1}{2}-\alpha\right)\Delta Q_{sc}+\frac{1}{2}\Delta Q_c  \end{equation}
For large $q_{sc}$ the head-tail modes with positive $k$ converge towards $Q_k=-\Delta Q_c/2$. For a given $k$ the mode is still inside the incoherent band if
 \begin{equation}
k\gtrsim\alpha q_{sc}-\frac{1}{2}q_c    
 \end{equation} holds. 
In order to illustrate the above analysis, the incoherent band for $k=2$ 
is shown in~\autoref{fig:ht-modes-band} (shaded area). For $q_c=0$ the coherent head-tail mode frequency 
crosses the upper boundary of the band at $q_{sc}\approx 4.5$. 
For $q_c=0.15q_{sc}$ the $k=2$ head-tail mode remains inside the incoherent band until $q_{sc}\approx 12$.    
For the $k=1$ modes the above analysis leads to thresholds of $q_{sc}\approx 2$ (thin beams) 
and $q_{sc}\approx 6$ for $q_c=0.15q_{sc}$.

\bild{85}{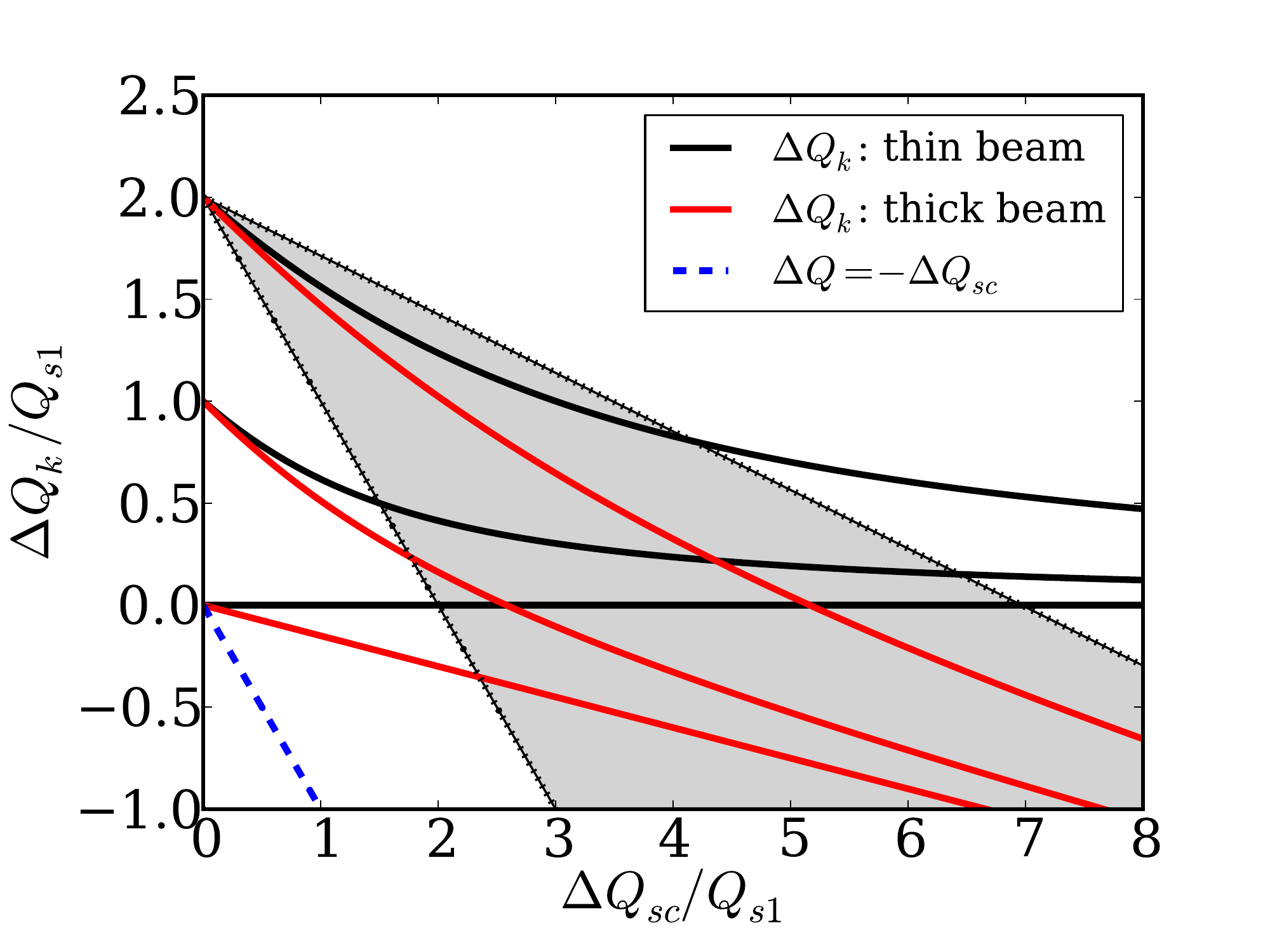}{fig:ht-modes-band}{Head-tail mode frequencies as a function of the space charge parameter. The grey shaded area indicates the incoherent band for a Gaussian bunch and $k=2$. 
The red curves represent the head-tail mode frequencies obtained for $q_c=0.15q_{sc}$.}

\section{Measurement setup for transverse bunch signals}\label{sec:meas-setup}

In this section, a brief description of the transverse beam excitation mechanisms as well as the two different tune measurement systems, TOPOS and BBQ, in the SIS-18 is given. Further the experimental set-up, typical beam parameters and uncertainty analysis of the measured beam parameters is discussed.

\subsection{Transverse Beam Excitation}
\begin{figure}[tbh]
   \centering
   \includegraphics*[width=85mm]{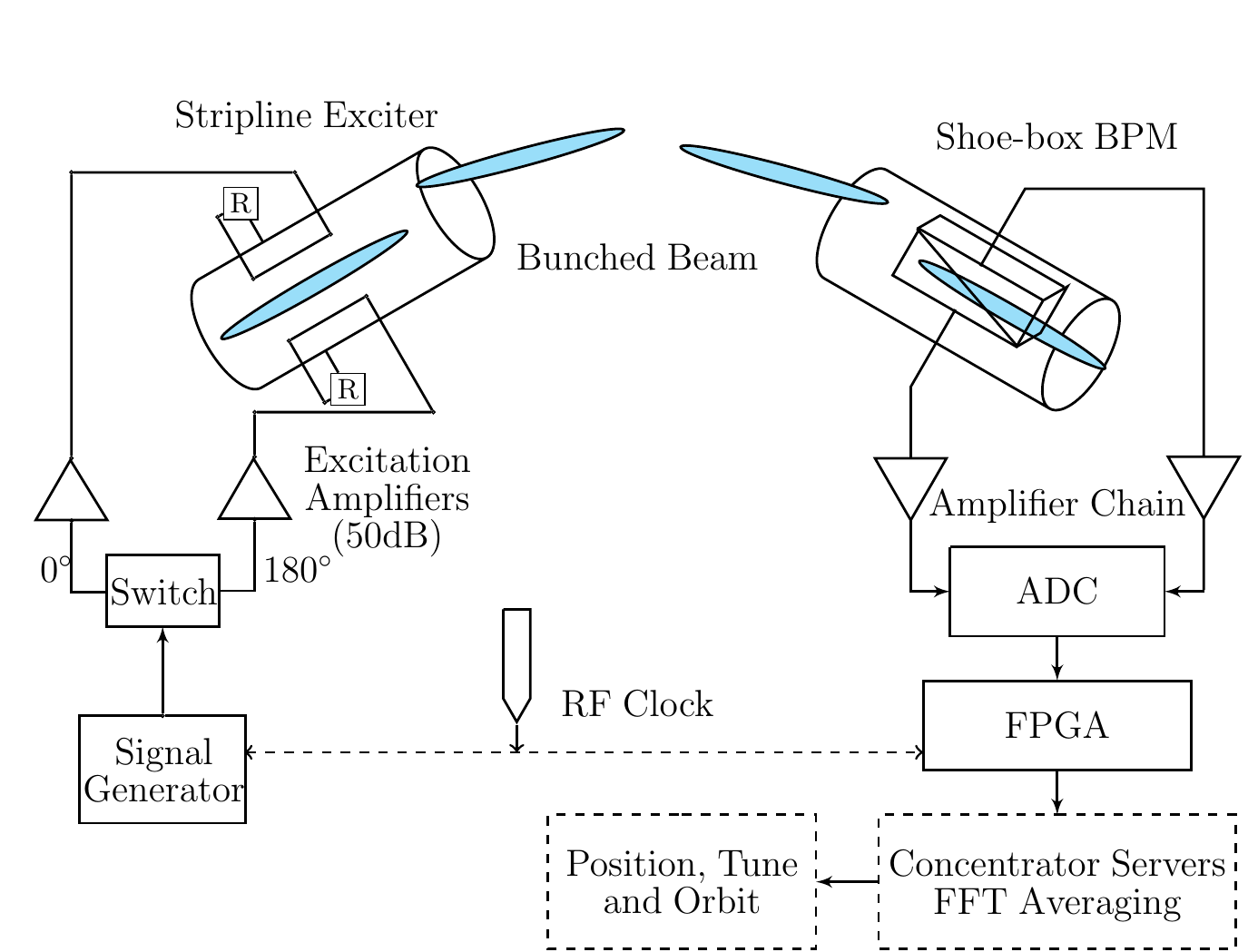}
   \caption{Scheme of the excitation system and the Tune, Orbit and Position (TOPOS) measurement system.}
   \label{fig:topos}
\end{figure}

The electronics used for beam excitation consist of a signal generator connected to two $25$ W amplifiers which feed power to $50 \Omega$ terminated stripline exciters as shown in~\autoref{fig:topos}. Excitation types such as band limited noise and frequency sweep are utilized at various power levels to induce coherent oscillations.

\subsubsection{Band limited noise:}
Band limited noise is a traditionally used beam excitation system for slow extraction in the SIS-18. The RF signal is mixed with Direct Digital Synthesis (DDS) generated fractional tune frequency, resulting in RF harmonics and their respective tune sidebands. This signal is further modulated by a pseudo-random sequence resulting in a finite band around the tune frequency. The width of this band is controlled by the frequency of the pseudo-random sequence. Typical bandwidth of band limited exciter is $\approx 5\%$ of the tune frequency. There are two main advantages of this system; first it is an easily tunable excitation source available during the whole acceleration ramp and second, the band limited nature of this noise results in an efficient excitation of the beam in comparison to white noise excitation. The main drawback is the difficulty in correlating the resultant tune spectrum with the excitation signal.

\subsubsection{Frequency sweep:}
Frequency sweep (chirp/harmonic excitation) using a network analyzer for BTF measurements is an established method primarily for beam stability analysis~\cite{Boussard1995}. However, using this method for tune measurements during acceleration is not trivial, and thus the method is not suitable for tune measurements during the whole ramp cycle. Nevertheless, this method offers advantages compared to the previous excitation method for careful interpretation of tune spectrum in storage mode, e.g., injection plateau or extraction flat top. Thus frequency sweep is used during measurements at injection plateau to compare and understand the dependence of tune spectra on the type of excitation.

\subsection{TOPOS}
Following the beam excitation, the signals from each of the 12 shoe-box type BPMs~\cite{Forck2004} at SIS-18 pass through a high dynamic range (90 dB) and broadband (100 MHz) amplifier chain from the synchrotron tunnel to the electronics room, where the signals are digitized using fast 14 bit ADCs at 125 MSa/s. Bunch-by-bunch position is calculated from these signals using FPGAs in real time and displayed in the control room. The spatial resolution is $\approx$0.5 mm in bunch-by-bunch mode. Further details can be found in \cite{Kowina2010}. Hence, TOPOS is a versatile system which provides accurate bunch-by-bunch position and longitudinal beam profile. This information is analyzed to extract non trivial parameters like betatron tune, synchrotron tune,  beam intensity evolution etc. 

\subsection{BBQ}
\begin{figure}[tbh]
   \centering
   \includegraphics*[width=100mm]{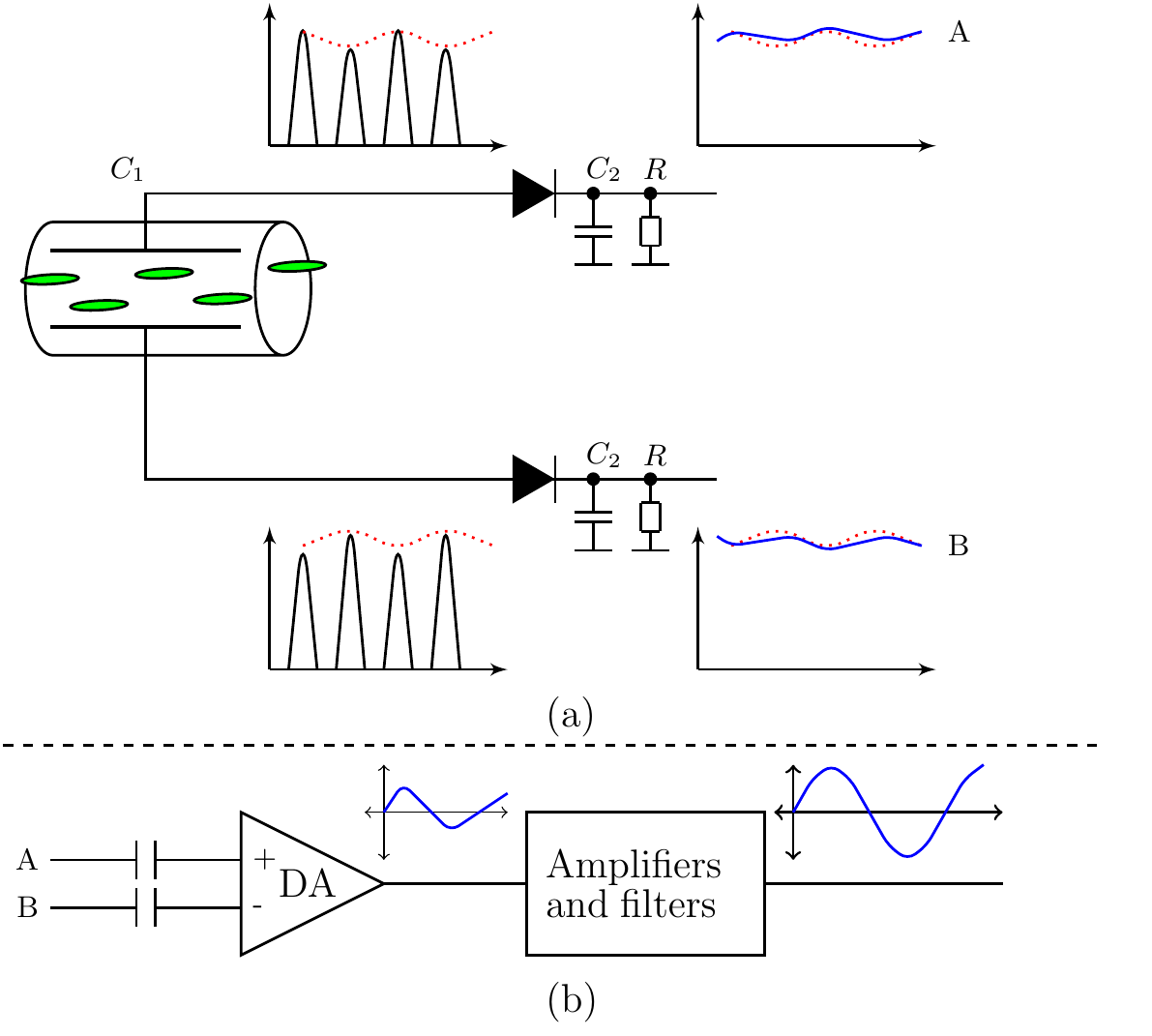}
   \caption{BBQ: Baseband Q measurement system. Diode detectors (a) and signal chain (b). }
   \label{fig:bbq}
\end{figure}

The BBQ system is a fully analog system and its front end  is divided into two distinct parts; a diode based peak detector 
and an analog signal processing chain consisting of input differential amplifier and a variable gain filter chain of 1 MHz bandwidth. The simple schematic of BBQ system configuration at SIS-18 is shown in~\autoref{fig:bbq} and the detailed principle of operation can be found in Ref. \cite{Gasior2012}.

\subsection{Comparison of TOPOS and BBQ}
The sensitivity of BBQ has been measured to be $\approx$10-15 dB higher than that of TOPOS under the present configuration. The main reason for the difference is the relative bandwidth of the two systems and their tune detection principles. In BBQ, the tune signal is obtained using analog electronics (diode based peak detectors and differential amplifier) immediately after the BPM plates, while in TOPOS position calculation is done after passing the whole bunch signal through a wide bandwidth amplifier chain. Even though the bunches are integrated to calculate position in TOPOS which serves as a low pass filter, the net signal-to-noise ratio is still below BBQ. Operations similar to BBQ could also be performed digitally in TOPOS to obtain higher sensitivity but would require higher computation and development costs. TOPOS can provide individual tune spectra of any of the four bunches in the machine, while BBQ system provides "averaged" tune spectra of all the bunches. Both the systems have been benchmarked against each other. Tune spectra shown in~\autoref{sec:ex-obser} are mostly from the BBQ system while TOPOS is primarily used for time domain analysis, nevertheless this will be pointed out when necessary.

\subsection{Beam parameters during the measurements}
\begin{figure}[tbh]
\centering
\includegraphics*[width=85mm]{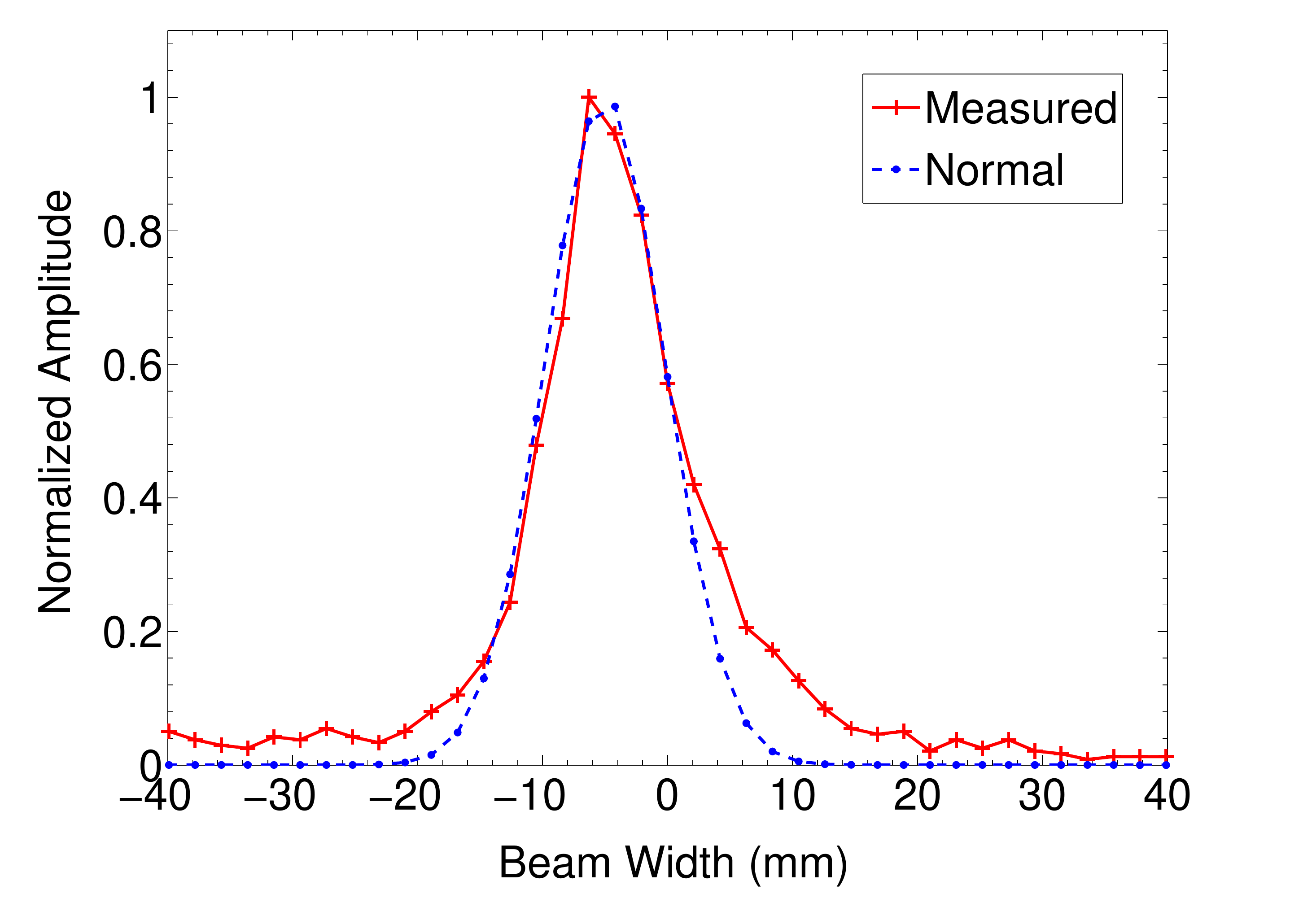}
\caption{Normalized transverse beam profile in vertical plane at $N^{7+} = 15\cdot10^8$ ions. The dotted lines shows the normal distribution for the rms width obtained by evaluation of beam profile around its centre.}
\label{fig:beam_profile}
\end{figure}

Experiments were carried out using N$^{7+}$ and U$^{73+}$ ion beams at the SIS-18 injection energy of 11.4\,MeV/u. 
The data were taken during 600\,ms long plateaus. At injection energy space charge effects are usually strongest. Four bunches are formed from the initially coasting beam during adiabatic RF capture. The experiment was repeated for different injection currents. At each intensity level several measurements were performed with different types and levels of beam excitation in both planes. Tune measurements were done simultaneously using the TOPOS and BBQ systems. The beam current and the transverse beam profile are measured using the beam current transformer~\cite{Reeg2001} and the ionization profile monitor (IPM)~\cite{Giacomini2004} respectively. An examplary transverse beam profile is shown in \autoref{fig:beam_profile}. The dipole synchrotron tune $(Q_{s1})$ is deduced using the residual longitudinal dipole fluctuations of the bunches. $Q_{s1}$ has been used as an effective synchrotron tune for all experimental results and will be referred as $Q_s$ from hereon. The momentum spread is obtained from longitudinal Schottky measurements~\cite{Caspers2008}. Important beam parameters during the experiment are given in~\autoref{tab:Beam Parameters U} and~\autoref{tab:Beam Parameters N}. It is important to note that all the parameters required for analytical determination of $q_{sc}, q_c$ are recorded during the experiments.
\begin{table}
\centering
\caption{Beam parameters during the $N^{7+}$ experiment}
\begin{tabular}{c c c}
\hline\hline
Beam/Machine parameters&Symbols & Values\\\hline
Atomic mass&$A$& 238\\\hline
Charge state&$q$ &73\\\hline
Kinetic energy &$E_{kin}$ & 11.4 MeV/u (measured\\\hline
Number of particles&$N_p$ & $1, 5, 12\cdot10^8$ (measured)\\\hline
Tune&$Q_{x},Q_{y}$ & 4.31, 3.27 (set value)\\\hline
Chromaticity&$\xi_{x},\xi_{y}$ & -0.94, -1.85 (set value)\\\hline
Transverse emittance&$\epsilon_x,\epsilon_y(2\sigma)$&$45, 22$ mm-mrad (measured)\\\hline
Slip factor&$\eta$& 0.94\\\hline
Bunching factor&$B_{f}$& 0.4 (measured)\\\hline
Synchrotron tune& $Q_{s0},Q_{s1}$ & 0.007,0.0065 (measured)\\\hline
Momentum spread& $\frac{\Delta p}{p}$ &0.001 (measured)\\\hline
Lattice parameter& $\beta_{x,trip}$, $\beta_{y,trip}$ & 5.49, 7.76\\\hline\hline
\end{tabular}
%\caption{Beam parameters during the $U^{73+}$experiment}
\label{tab:Beam Parameters U}
\end{table}

\begin{table}
\centering
\caption{Beam parameters during the $N^{7+}$ experiment}
\begin{tabular}{c c c}
\hline\hline
Beam/Machine Parameters & Symbols& Values\\\hline
Atomic mass& $A$ & 14\\\hline
Charge state&$q$ &7\\\hline
Kinetic energy&$E_{kin}$ & 11.56 MeV/u (measured)\\\hline
Number of particles&$N_p$ & $3, 6, 11, 15\cdot10^9$ (measured)\\\hline
Tune&$Q_{x},Q_{y}$ & 4.16, 3.27 (set value)\\\hline
Chromaticity&$\xi_{x},\xi_{y}$ & -0.94, -1.85 (set value)\\\hline
Transverse emittance&$\epsilon_x,\epsilon_y(2\sigma)$&$33, 12$ mm-mrad (measured)\\\hline
Slip factor&$\eta$& 0.94\\\hline
Bunching factor&$B_{f}$& 0.37 (measured)\\\hline
Synchrotron tune&$Q_{s0},Q_{s1}$   & 0.006,0.0057 (measured)\\\hline
Momentum spread&$\frac{\Delta p}{p}$  &0.0015 (measured)\\\hline
Lattice parameter&$\beta_{x,trip}$, $\beta_{y,trip}$ & 5.49, 7.76\\\hline\hline
\end{tabular}
\label{tab:Beam Parameters N}
\end{table}

From \autoref{tab:Beam Parameters U} and \autoref{tab:Beam Parameters N} one can estimate
that in the measurements the head-tail space charge and image current parameters
were in the range $q_{sc}\lesssim 10$ and $q_c\lesssim 0.2 q_{sc}$ for the 
horizontal and vertical planes. 

\section{Experimental Observations}\label{sec:ex-obser}
Tune spectra measurements at different currents are presented and interpreted in comparison with the predictions of~\autoref{sec:ht-modes}. The effect of different excitation types and power on the transverse tune spectra is studied. Transverse impedances in both planes are obtained from the coherent tune shifts. Incoherent tune shifts are obtained from the relative frequency shift of head-tail modes in accordance with~\autoref{eq:htmodes}. Chromaticity is measured using head-tail eigenmodes and the obtained relative height of the observed peaks for different chromaticities is analyzed. 
\subsection{Modification of the tune spectrum with intensity}\label{subsec:tun_int}
\begin{figure}[tbh]
\centering
   \includegraphics*[width=85mm]{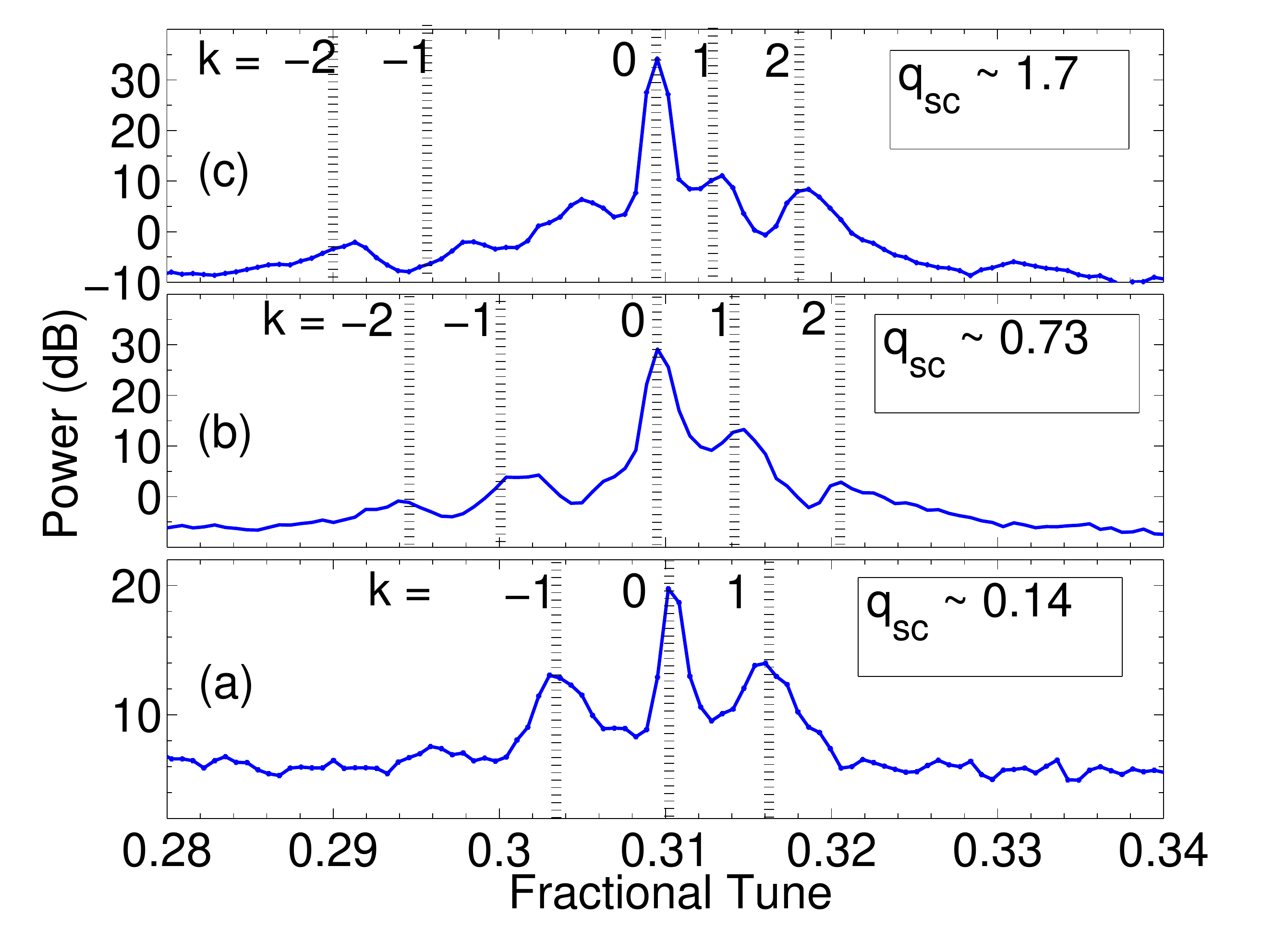}
   \caption{Horizontal tune spectra for $U^{73+}$ ions and beam parameters given in~\autoref{tab:Beam Parameters U} (see text). The dashed lines indicate the head-tail tune shifts from~\autoref{eq:htmodes}.}
   \label{hor_spectra_U73}
\end{figure}

\Autoref{hor_spectra_U73} shows the horizontal tune spectra obtained with the BBQ system using band-width limited noise at different intensities.~\Autoref{hor_spectra_U73}(a) shows the horizontal tune spectrum at low intensity. Here the $k=1,0,-1$ peaks are almost equidistant, which is expected for low intensity bunches. The space charge parameter obtained using the beam parameters and~\autoref{eq:sctune} is $q_{sc}\approx 0.15$. The vertical lines indicate the positions of the synchrotron satellites obtained from~\autoref{eq:htmodes} (with $Q_s=Q_{s1}$). 
\Autoref{hor_spectra_U73}(b) shows the tune spectrum at moderate intensity ($q_{sc}\approx 0.7$). 
The $k=2,-2$ peaks can both still be identified.~\Autoref{hor_spectra_U73}(c) shows the tune spectra at larger intensity ($q_{sc}\approx 1.7$). An additional peak appears between the $k=0$ and $k=-1$ peaks which can be attributed to the mixing product of diode detectors (since at this intensity $30-40 V$ acts across the  diodes pushing it into the non-linear regime). The $k=0,1,2$ peaks can be identified very well, whereas the amplitudes of the lines for negative $k$ already start to decrease (see~\autoref{sec:ht-modes}). In the horizontal plane the effect of the pipe impedance 
and the corresponding coherent tune shift can usually be neglected because of 
the larger pipe diameter.

\begin{figure}[tbh]
\centering
   \includegraphics*[width=85mm]{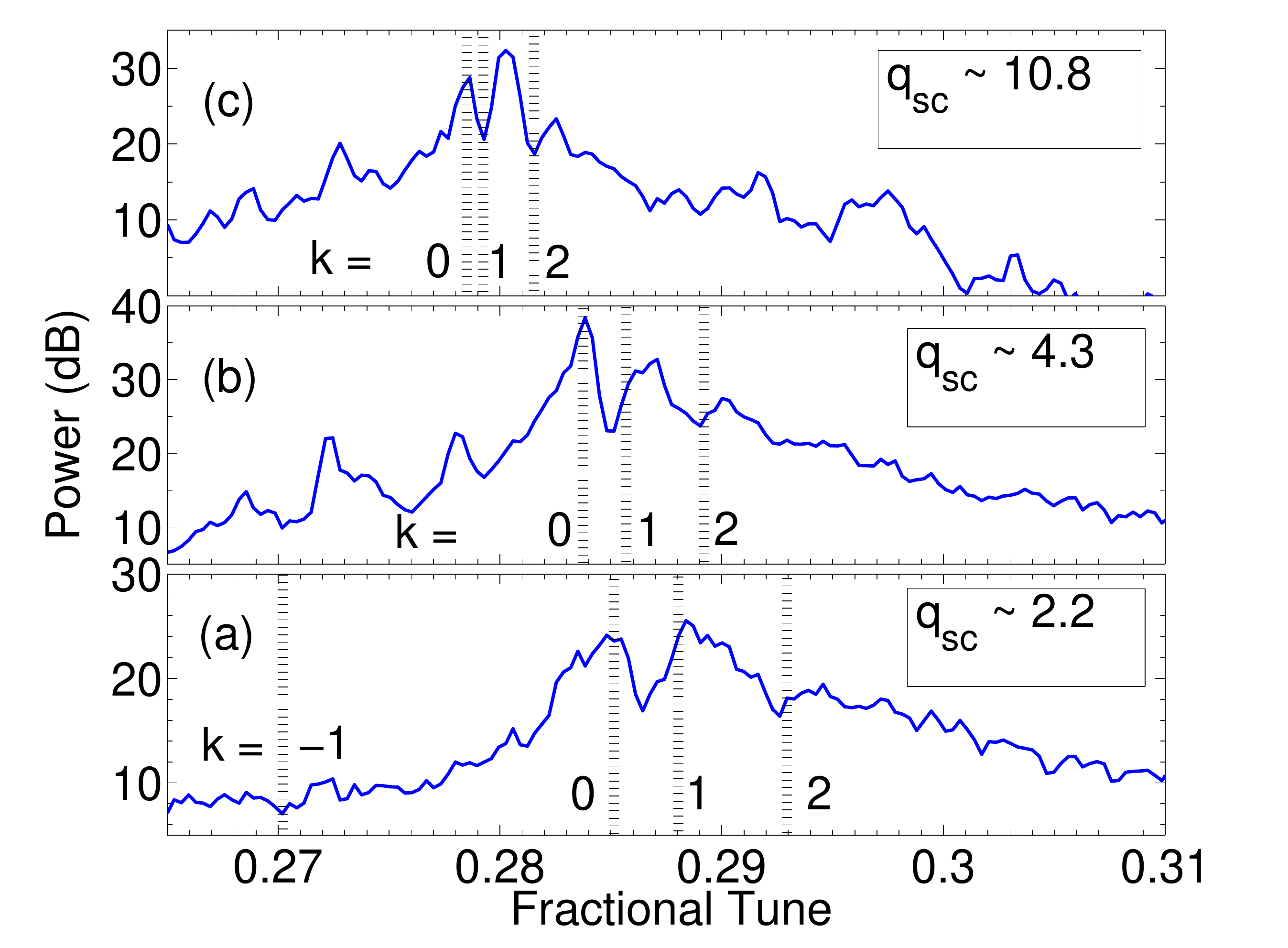}
   \caption{Vertical tune spectra for $N^{7+}$ ions and beam parameters given in~\autoref{tab:Beam Parameters N}. The dashed lines indicate the head-tail tune shifts from~\autoref{eq:htmodes}.}
   \label{ver_spectra_N7}
\end{figure}

\Autoref{ver_spectra_N7} shows the vertical tune spectrum obtained by the BBQ system with band limited noise excitation for $N^{7+}$ beams with $q_{sc}$ values larger than $2$. Here the negative modes $(k<0)$ could not be resolved anymore. In the vertical plane the coherent tune shift is larger due to the smaller SIS-18 beam pipe diameter ($q_{c}\approx q_{sc}/10$). The shift of the $k=0$ peak
due to the effect of the pipe impedance is clearly visible in~\autoref{ver_spectra_N7}.

In the measurements the width of the peaks is determined by the cumulative effect of non-linear synchrotron motion, non-linearities of the optical elements, closed orbit distortion, tune fluctuation during the measurement interval as well as due to the intrinsic Landau damping (\autoref{sec:ht-modes}). From the comparison 
to the simulations we conclude that the intrinsic Landau damping is an important contribution
to the width of the $k=1,2$ peaks. 

\subsection{Determination of coherent and incoherent tune shifts}\label{subsec:inc_tune_shift}
\begin{figure}[tbh]
\centering
\includegraphics*[width=85mm]{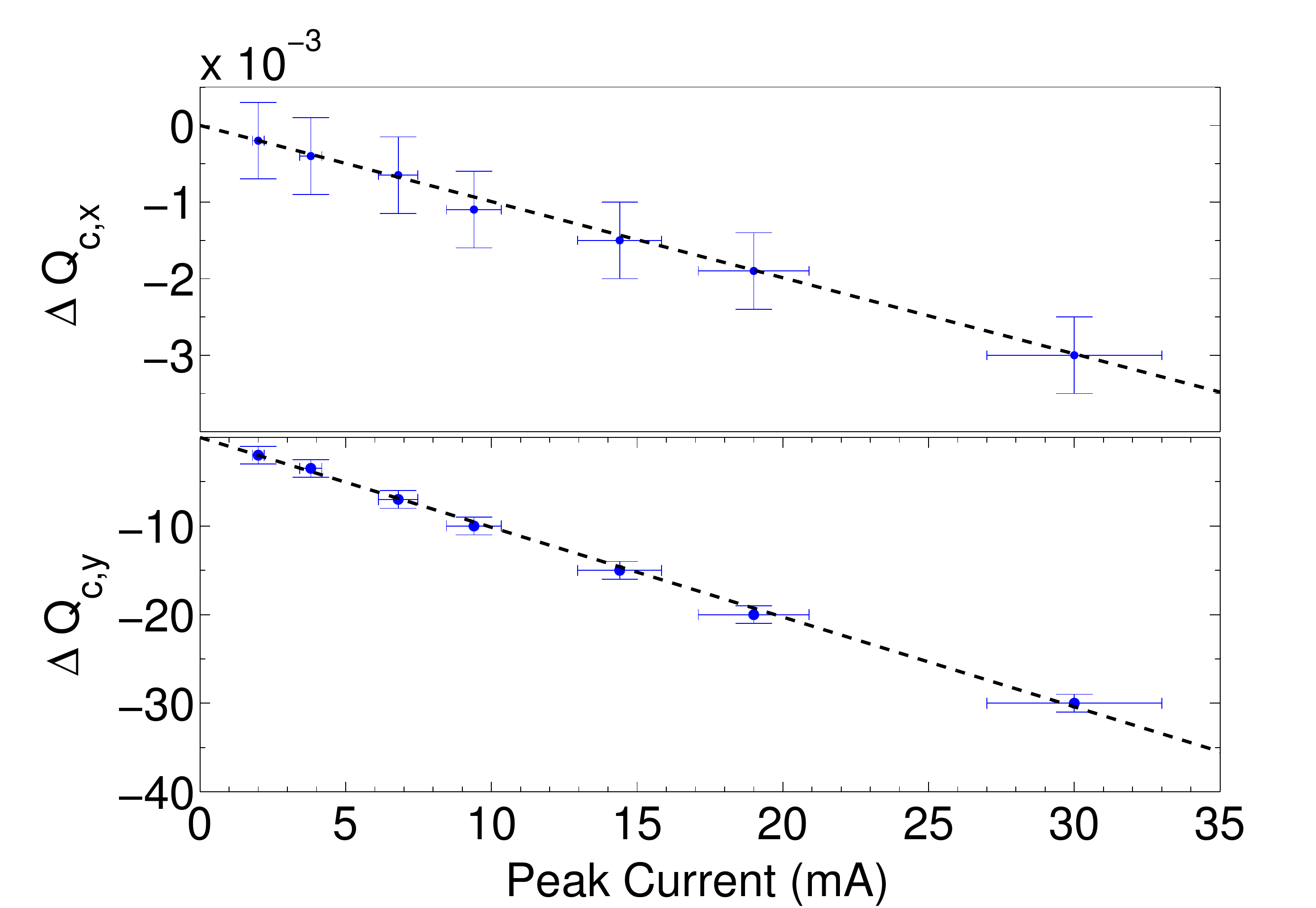}
\caption{Coherent tune shift obtained from the measurement for the horizontal and for the vertical planes as a function of the peak beam current. The dotted lines correspond to a linear least square error fit. The error bars in the horizontal plane are due to uncertainties in the current measurements. In the vertical plane errors result from the width of $k=0$ mode. The FFT resolution is $\approx 5\cdot 10^{-4}$ and is always kept higher than the mode width.}
\label{coh_shift}
\end{figure}

The coherent tune shift $\Delta Q_c$ can be obtained by measuring shift of the $k=0$ line as a function of the 
peak bunch current as shown in~\autoref{coh_shift}. The transverse impedance is obtained by a linear least square error fit of the measured shifts in both planes to~\autoref{eq:impedance} and~\autoref{eq:coh_shift}. The impedance values are
obtained in the horizontal and vertical planes at injection energy are $0.23$ M$\Omega/m^2$ and $1.8$ M$\Omega/m^2$
respectively, which agrees very well with the expected values for the average beam pipe radii of the SIS-18.  

\begin{figure}[tbh]
\centering
\includegraphics*[width=85mm]{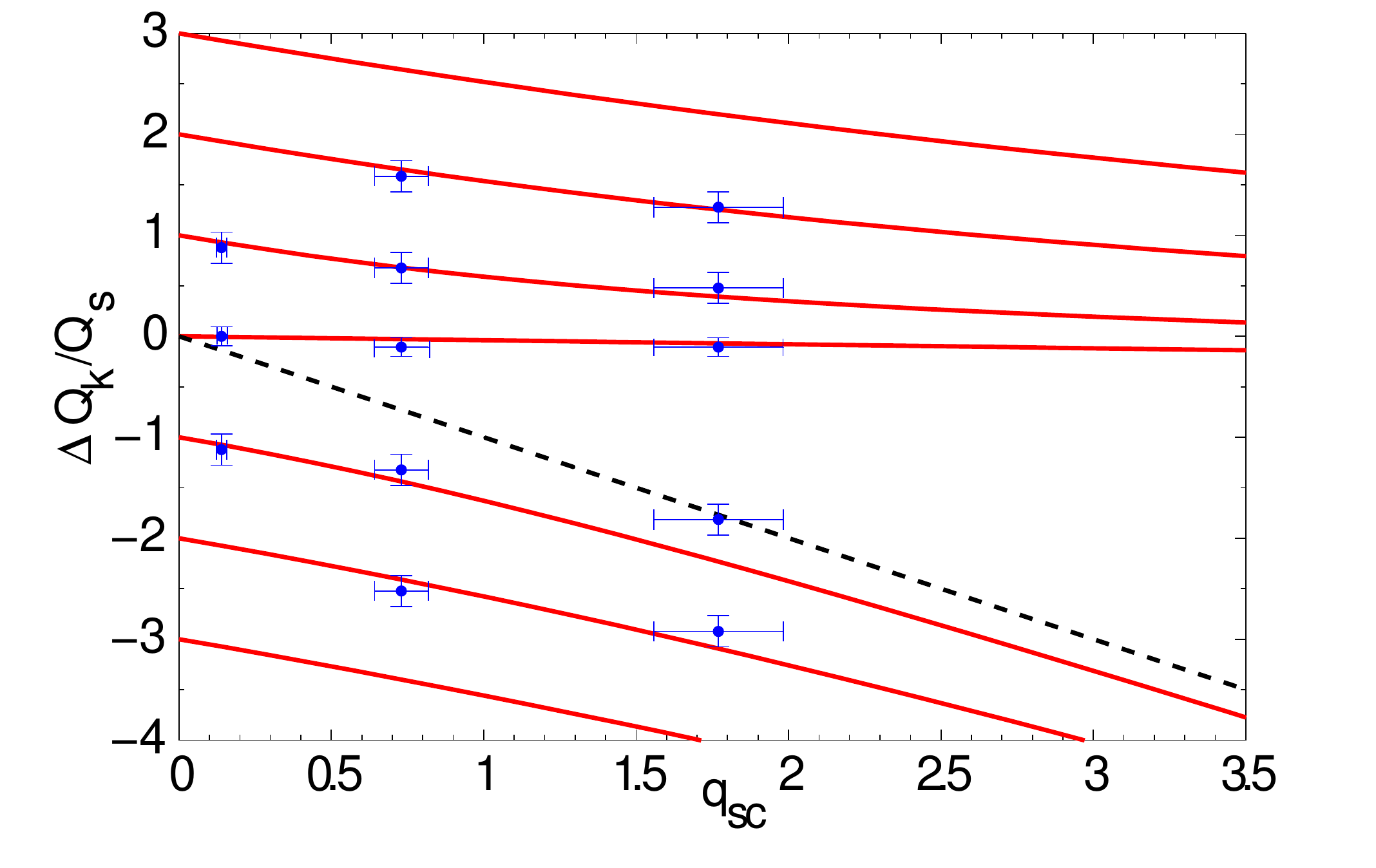}
\caption{The measured positions of the peaks in the horizontal tune spectra for different $U^{73+}$ 
beam intensities together with the analytical curves from~\autoref{eq:htmodes} using the space charge tune shift estimated from the beam parameters in~\autoref{tab:Beam Parameters U}. The dotted line corresponds to the incoherent tune shift. The error bars for the vertical plane correspond to the width of measured modes (see text). In the horizontal plane the error bars are estimated by the propagation of uncertainties (see text).}
\label{fig:inc_shift1}
\end{figure}
\begin{figure}[tbh]
\centering
\includegraphics*[width=85mm]{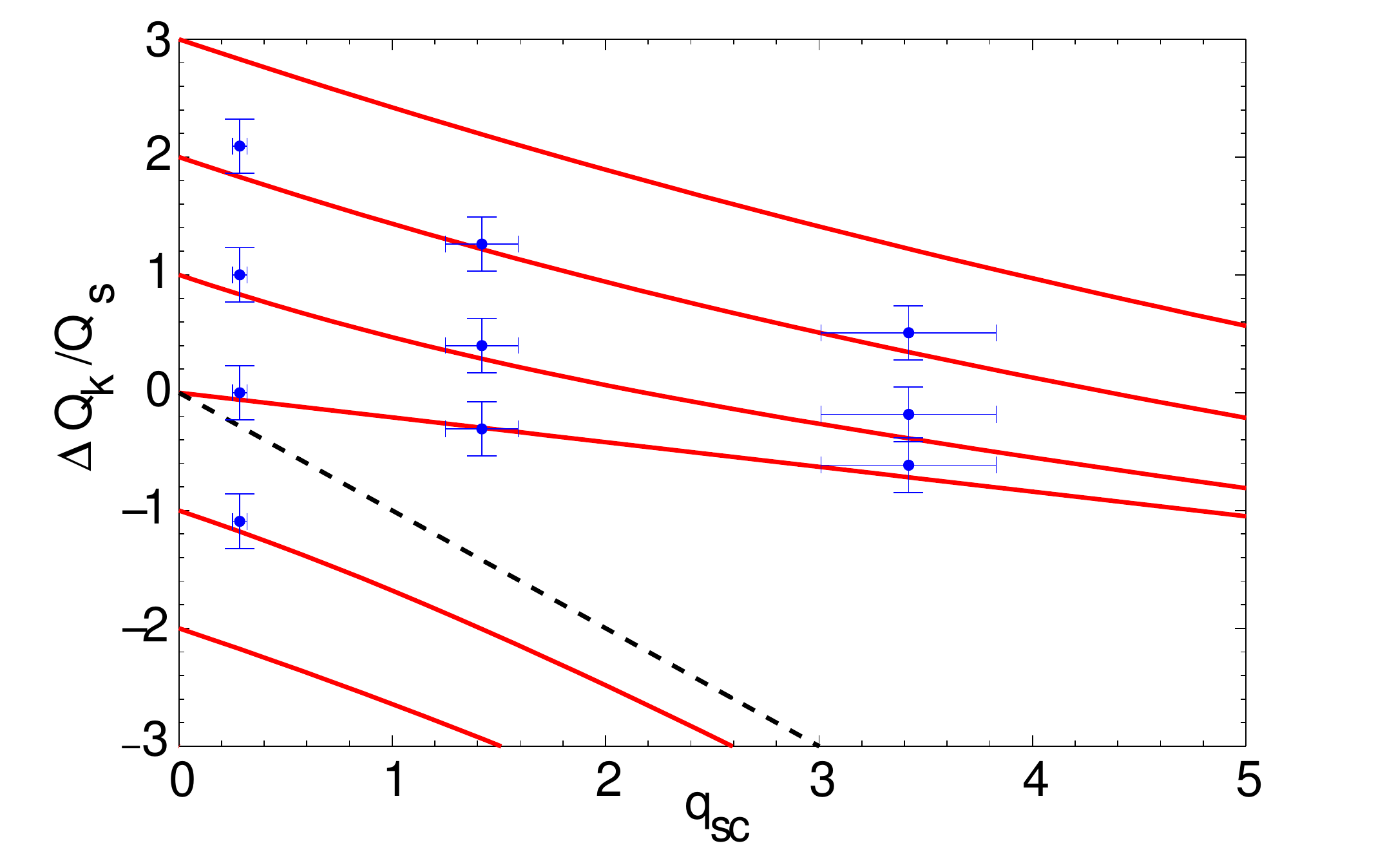}
\caption{This plot shows the predicted shifts from analytical~\autoref{eq:htmodes} and measured head-tail mode frequencies are overlaid  in vertical plane for $U^{73+}$ ion beam at various current levels (\autoref{tab:Beam Parameters U}).}
\label{fig:inc_shift2}
\end{figure}
\begin{figure}[tbh]
\centering
\includegraphics*[width=85mm]{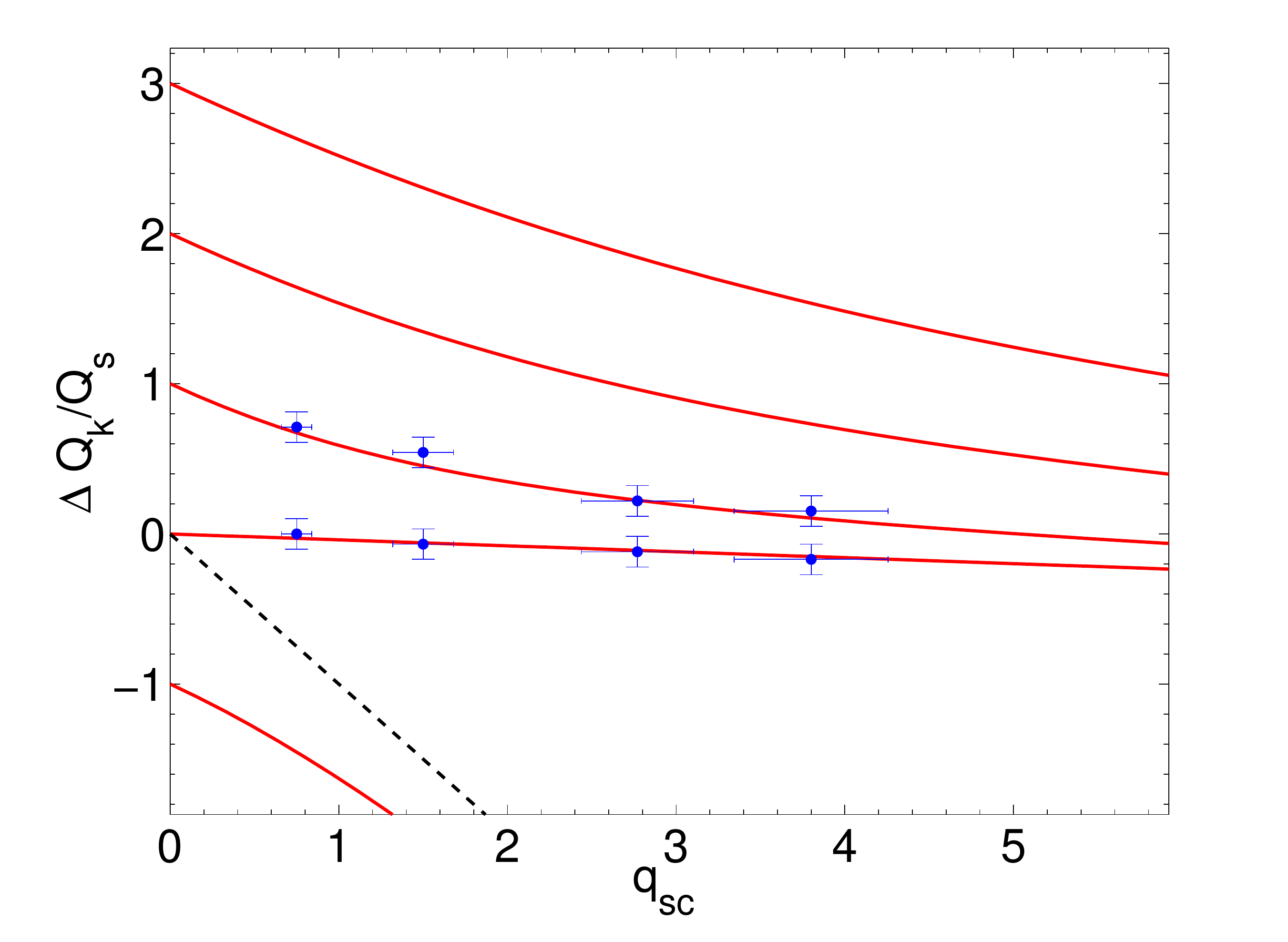}
\caption{This plot shows the predicted shifts from analytical~\autoref{eq:htmodes} and measured head-tail mode frequencies are overlaid  in horizontal plane for $N^{7+}$ ion beam at various current levels (\autoref{tab:Beam Parameters N}).}
\label{fig:inc_shift3}
\end{figure}
\begin{figure}[tbh]
\centering
\includegraphics*[width=85mm]{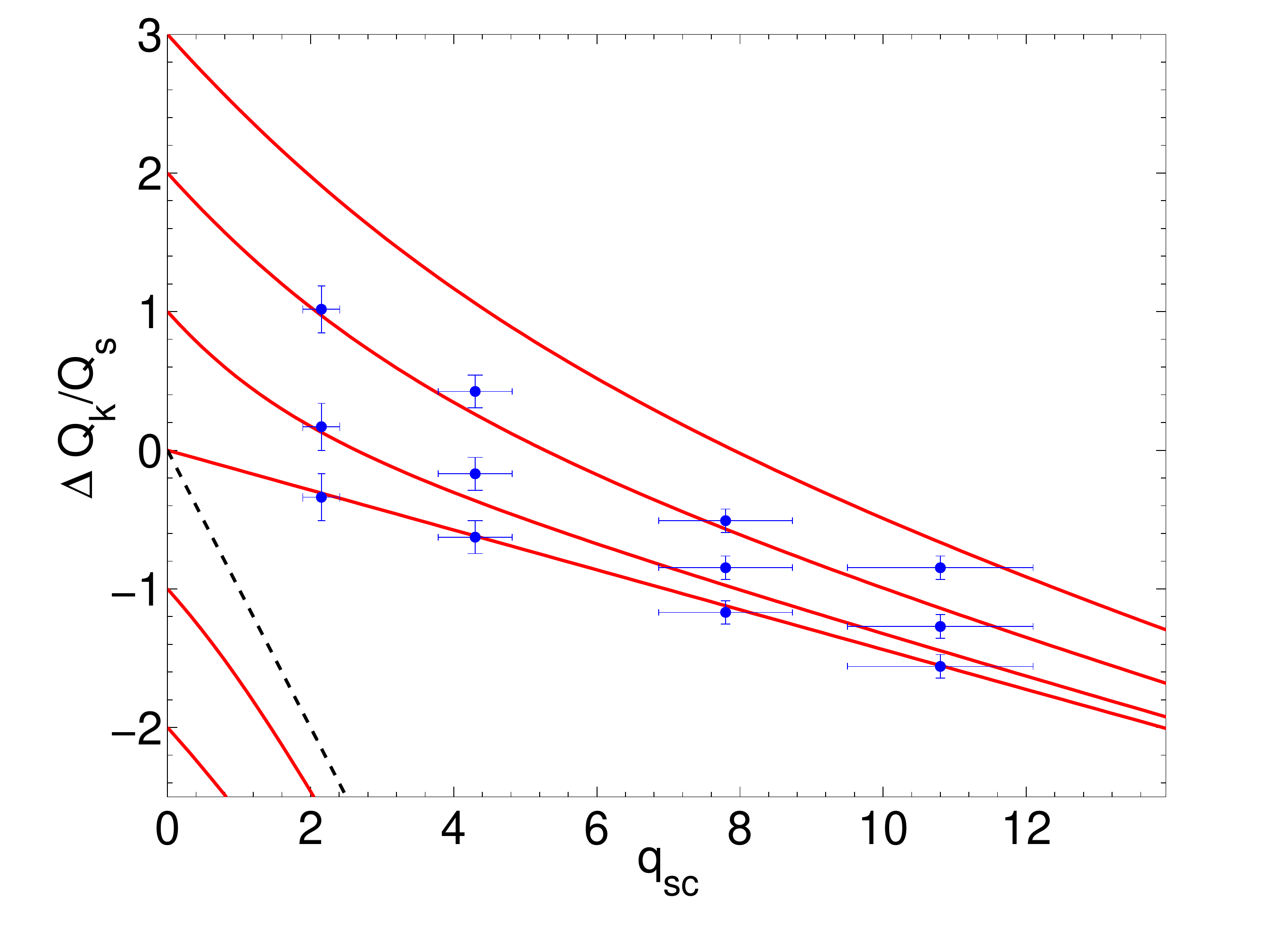}
\caption{This plot shows the predicted shifts from analytical~\autoref{eq:htmodes} and measured head-tail mode frequencies are overlaid  in vertical plane for $N^{7+}$ ion beam at various current levels (\autoref{tab:Beam Parameters N}).}
\label{fig:inc_shift4}
\end{figure}

Figures~\ref{fig:inc_shift1},~\ref{fig:inc_shift2},~\ref{fig:inc_shift3} and~\ref{fig:inc_shift4} show the measured 
positions of the peaks in the tune spectra for different intensities. In comparison the analytical curves (solid lines) obtained from~\autoref{eq:htmodes} for the head-tail tune shifts are plotted using $q_{sc}$ estimated from the beam parameters in~\autoref{tab:Beam Parameters U} and~\autoref{tab:Beam Parameters N} for each intensity. The error bars in the vertical plane ($\delta q_k = \frac{\Delta Q_{k}}{Q_s}$) correspond to the $3$ dB width of the measured peak due to accumulation of various effects (see~\autoref{subsec:tun_int}). In the horizontal plane, error bars ($\delta q_{sc}$) are estimated by propagation of parameter uncertainties mentioned in~\autoref{subsec:meas_err}. 

In this subsection, we introduce another space charge parameter $q_{sc,m}$ which is the measured space charge parameter using the following method. It is not to be confused with $q_{sc}$ which is predicted for a given set of beam parameters by~\autoref{eq:sctune}. The incoherent space charge tune shift can be determined directly from the tune spectra by measuring the separation between the $k=0$ and $k=1$ peaks, i.e. $ (q_{k01} = \frac{\Delta Q_{k01}}{Q_s}$) and fitting it with the parameter $q_{sc}$ in the predictions from~\autoref{eq:htmodes}. The value of $q_{sc}$ for the best fit is denoted as $q_{sc,m}$.

\begin{align}\label{eq:sc_meas_error}
&q_{sc,m} = \frac{1 - q_{k01}^2}{\gamma q_{k01}} \hspace{5cm}  0\leq q_{k01} \leq 1 
\end{align}
\Autoref{eq:sc_meas_error} is obtained by rearranging~\autoref{eq:htmodes} for $k= 0,1$ while  $\gamma = \frac{ q_{sc}-q_c}{q_{sc}}$. The linearized absolute error on measured $q_{sc,m}$ ($\delta q_{sc,m}$) is given by 

\begin{align}
&\delta q_{sc,m} = \frac{-(1+ q_{k01}^2)}{\gamma q_{k01}^2}\cdot\delta q_{k01}
\end{align}
 $\delta q_{k01}$ is given by either the width of the $k =0,1$ lines or by the frequency resolution of the system.  In a typical tune spectra measurement using data from 4000 turns $\delta q_{k01}\approx 0.04$. The absolute error is a non-linear function of $q_{sc}$ in accordance to the~\autoref{eq:sc_meas_error}.
It is possible to define the upper limit of $q_{sc}$ where this method is still adequate based on the system resolution and~\autoref{eq:sc_meas_error}. If we define a criterion that,  $ q_{k01} \gtrsim \delta q_{k01}$ to resolve the head-tail modes. This gives the limit to be $q_{sc} \lesssim 8$ where the measurement error is still within the defined criterion.

\begin{figure}[tbh]
\centering
\includegraphics*[width=85mm]{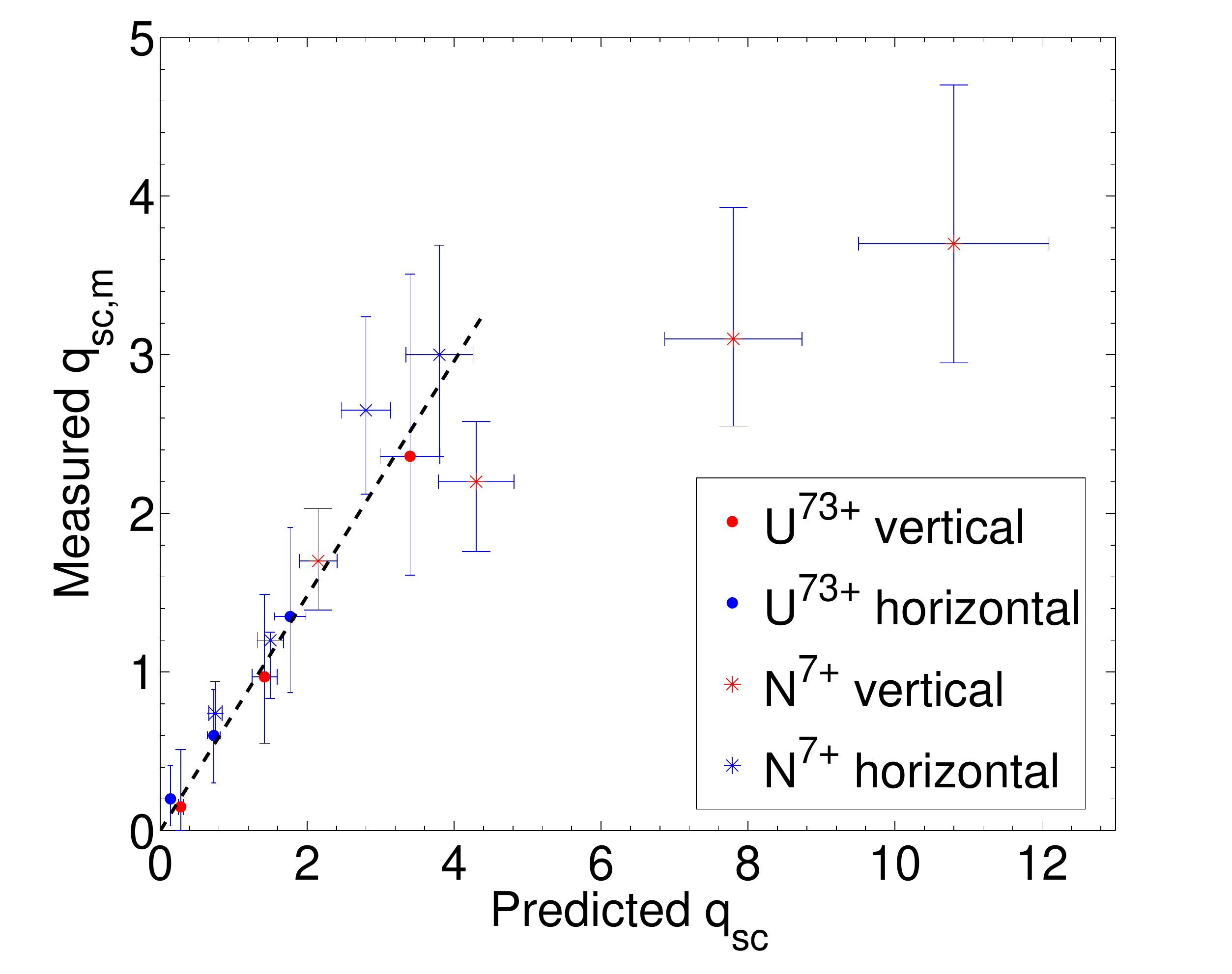}
\caption{Combining the results from the Figures~\ref{fig:inc_shift1},~\ref{fig:inc_shift2},~\ref{fig:inc_shift3} and~\ref{fig:inc_shift4}, a plot of predicted $q_{sc}$ using~\autoref{eq:sctune} against measured $q_{sc}$ using the distance between modes $k=0$ and $k=1$ is obtained.}
\label{fig:pred_meas}
\end{figure}

 \Autoref{fig:pred_meas} shows a plot of the predicted space charge tune shifts ($q_{sc}$) versus the ones measured from the tune spectra using the above procedure ($q_{sc,m}$). For $q_{sc}\lesssim 3.5$ the space charge tune shifts measured from the tune spectra are systematically
lower by a factor $= 0.74$ than the predicted shifts. It is shown by the dotted line in~\autoref{fig:pred_meas} which is obtained by total least squares fit of the measured data points. 
For larger $q_{sc}$ the factor decreases to $\approx 0.4$.  Thus the method for measuring the incoherent tune shift based on head-tail tune shifts is found to be satisfactory only in the range $q_{sc}\lesssim 3.5$.  A possible explanation is the effect of the pipe impedance. Similar observations are made by the results of self-consistent simulations in~\autoref{sec:ht-modes}, where for $q_{sc}\gtrsim 2$ the separation of the $k=0$ and $k=1$ peaks observed is underestimated by~\autoref{eq:htmodes} (see~\autoref{fig:patric_thick}).

\subsection{Effect of excitation parameters on tune spectrum}
\begin{figure}[tbh]
\centering
\includegraphics*[width=85mm]{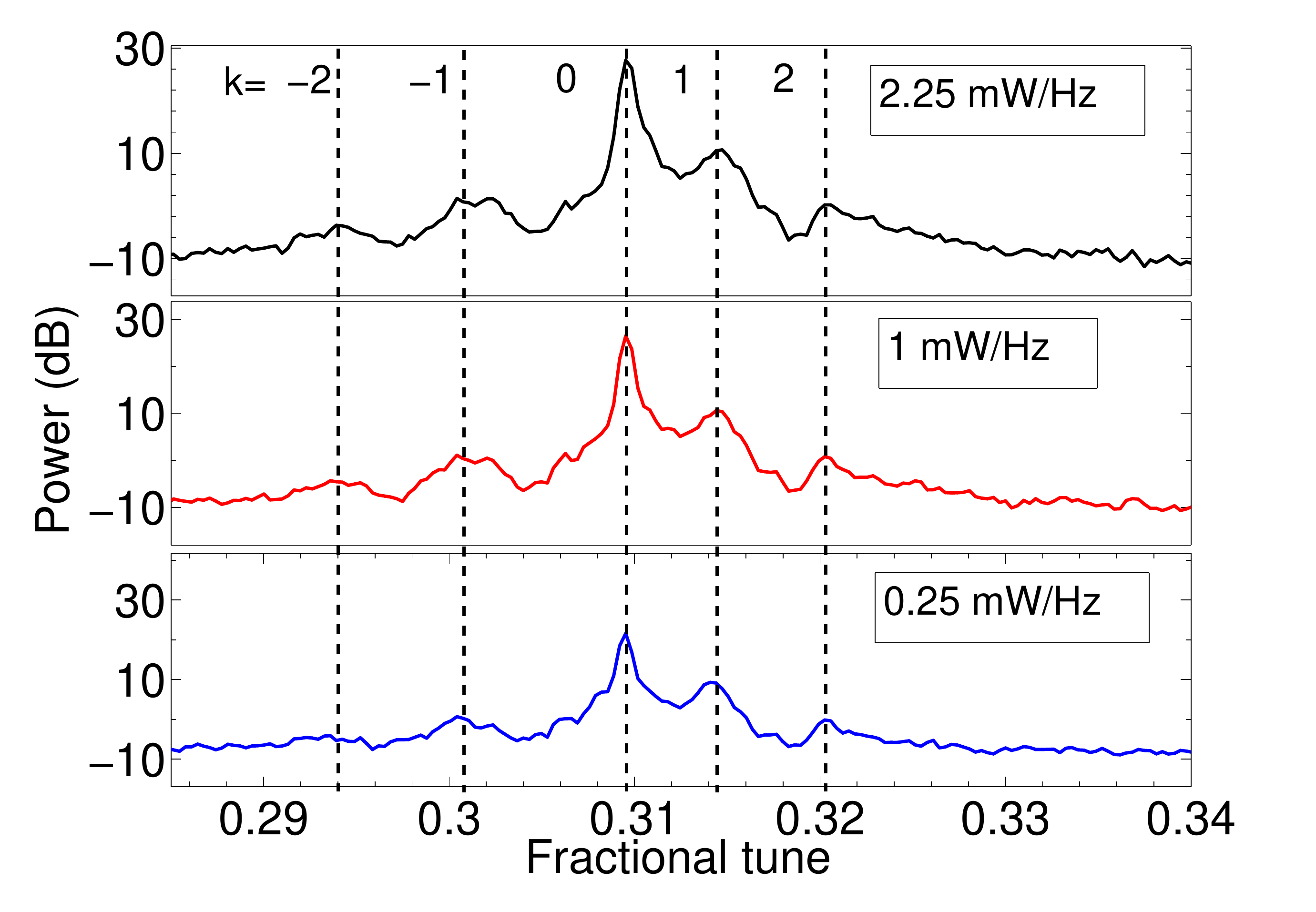}
\caption{Tune spectra at $5\cdot10^8 U^{73+}$ particles with beam excited at three different excitation amplitudes. The beam is excited with $0.25,1.0$ and $2.25$ mW/Hz power (blue, red and black respectively). The dotted lines mark the relative positions of the head-tail modes. The frequencies of various head-tail modes are unaffected.}
\label{fig:excitation_power}
\end{figure}

\Autoref{fig:excitation_power} presents the tune spectra obtained from BBQ system at various excitation power levels of band limited noise. The beam is excited with $0.25,1.0$ and $2.25$ mW/Hz power spectral density on a bandwidth of $10$ KHz. Signal-to-noise ratio (SNR) increases with excitation power whereas the spectral position of various modes is independent of excitation power. 

\begin{figure}[tbh]
\centering
   \includegraphics*[width=85mm]{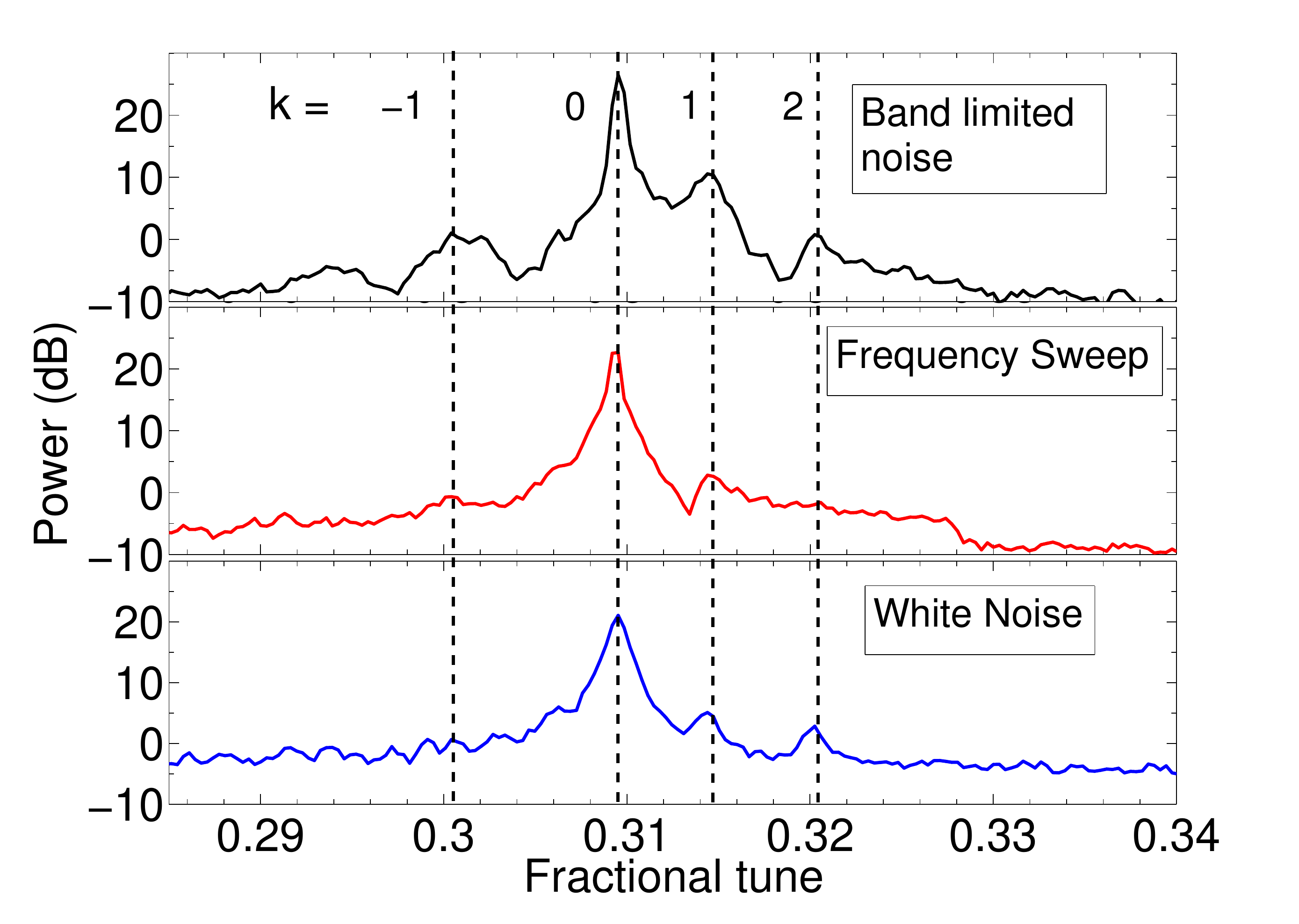}
   \caption{Horizontal tune spectra over time with $U^{73+}$ at $5\cdot10^8$ particles with band limited noise (RF Noise), frequency sweep (chirp noise) and white noise (top to bottom). The dotted lines mark the relative positions of the head-tail modes.}
   \label{fig:excitation_type-compare-fig}
\end{figure}

Beam excitation using two other excitation types i.e. frequency sweep and white noise is also performed to study the effect of excitation type on the tune spectra.~\Autoref{fig:excitation_type-compare-fig} shows the tune spectra under same beam conditions for different types of beam excitation obtained from the BBQ system. The frequencies of various modes in the tune spectra are independent of the type of excitation. The signal-to-noise ratio (SNR) is optimum for band limited noise due to long averaging time compared to "one shot" spectra from sweep excitation.

\subsection{Time domain identification of head tail modes}
\begin{figure}[tbh]
\centering
   \includegraphics*[width=85mm]{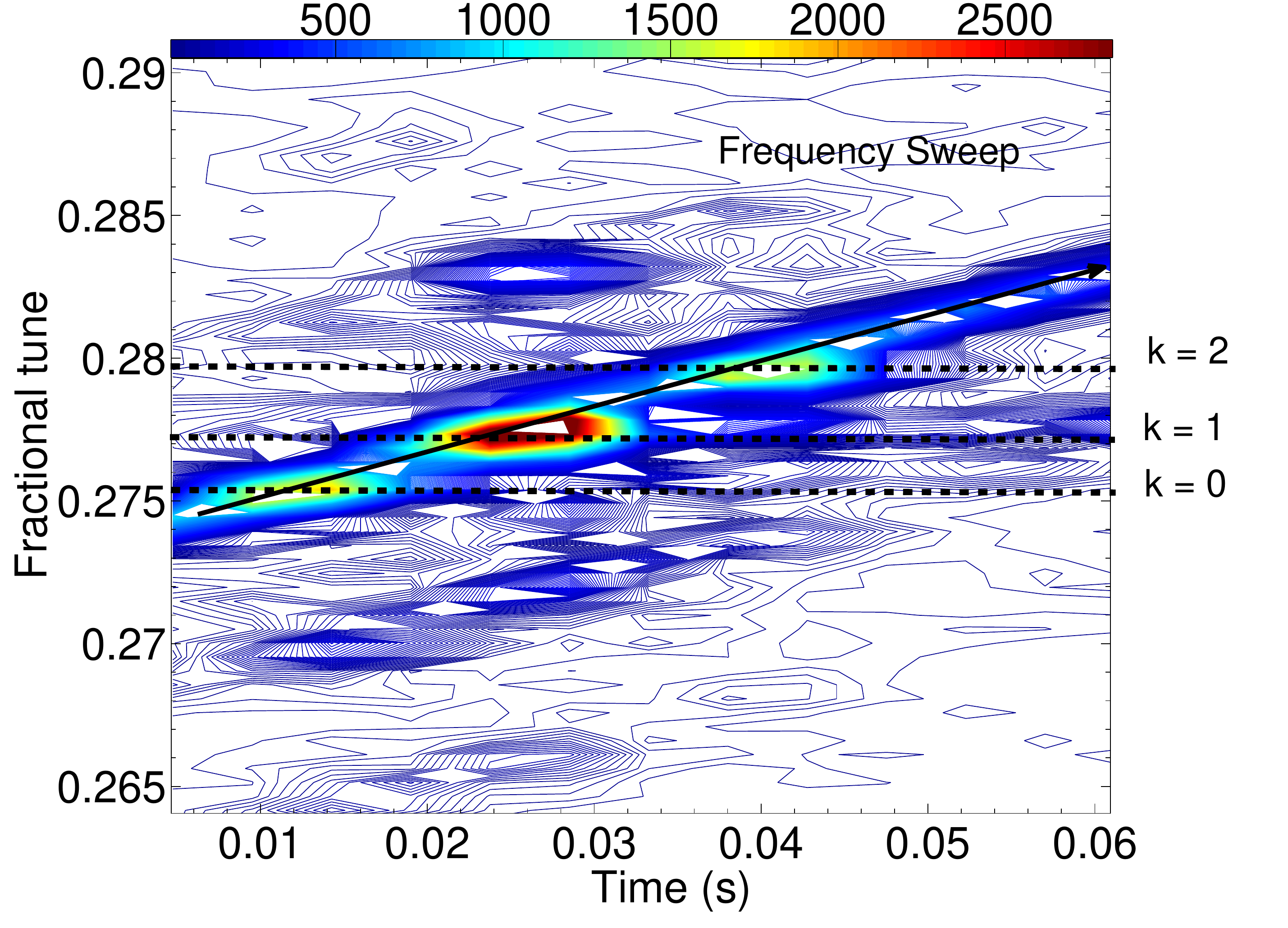}
   \caption{Vertical tune spectra with frequency sweep over time at  $N^{7+}$ with $1.5\cdot10^9$ particles. The various modes get individually excited as the sweep frequency (marked by the arrow) coincides with the mode frequency. The $k=0, 1, 2$ modes are marked. Symmetric sidebands due to coherent synchrotron oscillations are visible around the sweep frequency spaced with $Q_s$ and are clearly distinguished from the head-tail modes.}
   \label{fig:chirp-tune-fig}
\end{figure}

\begin{figure}[tbh]
\centering
   \includegraphics*[width=85mm]{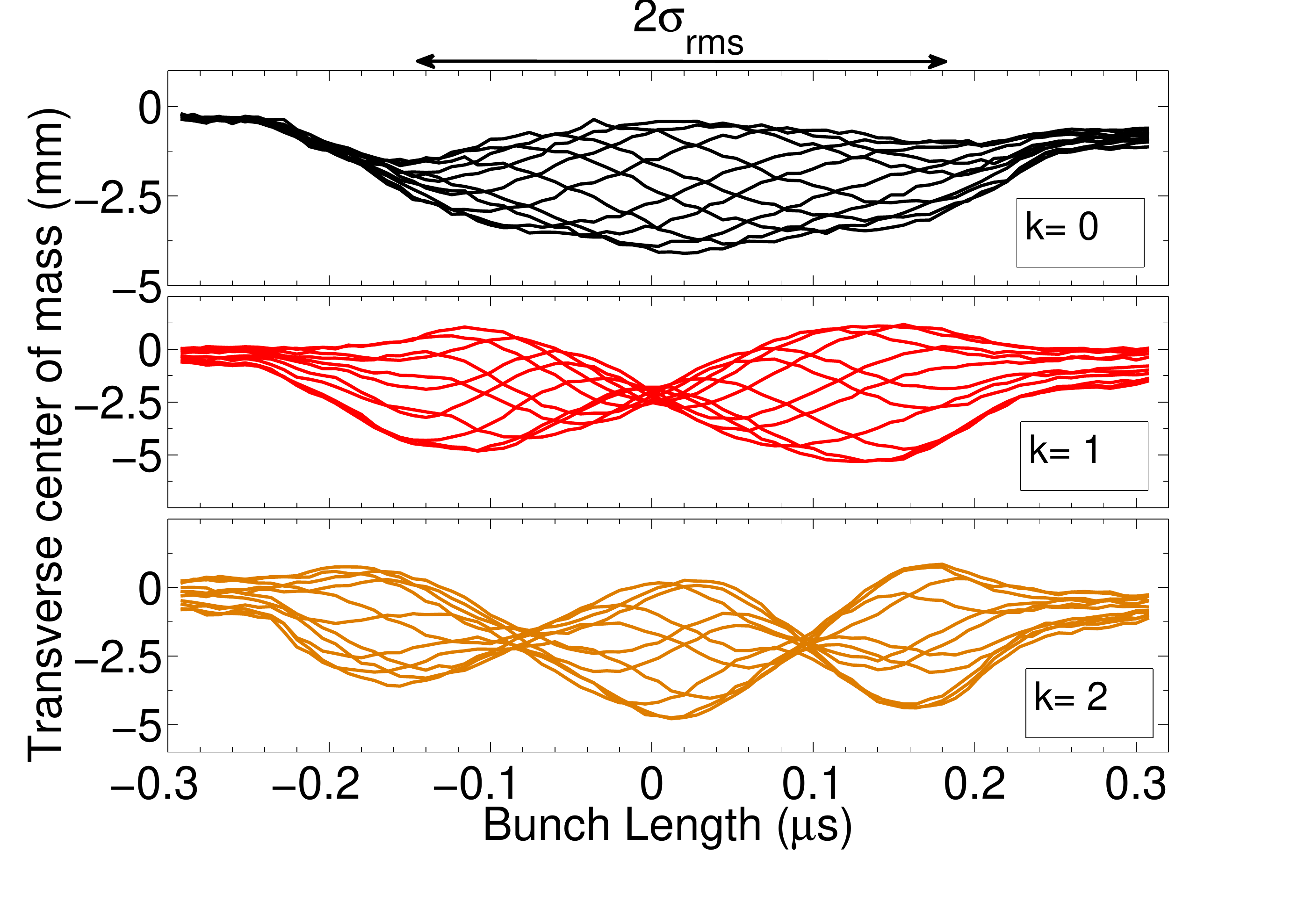}
   \caption{ Transverse center of mass along the bunch for $k=0, 1$ and $2$ modes corresponding to~\autoref{fig:chirp-tune-fig}. Frequency sweep is used to excite the beam which enables to resolve the distinct head-tail modes temporally. The rms bunch length ($2\sigma_{rms}$) is indicated.}
   \label{fig:mode-fig}
\end{figure}

\Autoref{fig:chirp-tune-fig} shows the 2-D contour plot for frequency sweep excitation in vertical plane obtained from the TOPOS system, where various head-tail modes are individually excited as the excitation frequency crosses them. Frequency sweep excitation allows resolving the transverse center of mass along the bunch for various modes which helps in identifying each head-tail mode in time domain. This serves as a direct cross-check for the spectral information and leaves no ambiguity in identification of the order (k) of the modes. \Autoref{fig:mode-fig} shows the corresponding transverse center of mass along the bunch for $k=0,1$ and $2$ at the excited time instances. This method works only with sweep excitation and requires high signal-to-noise ratio in the time domain, which amounts to higher beam current or high excitation power.

\subsection{Measurement of the chromaticity}

\begin{figure}[tbh]
\centering
   \includegraphics*[width=85mm]{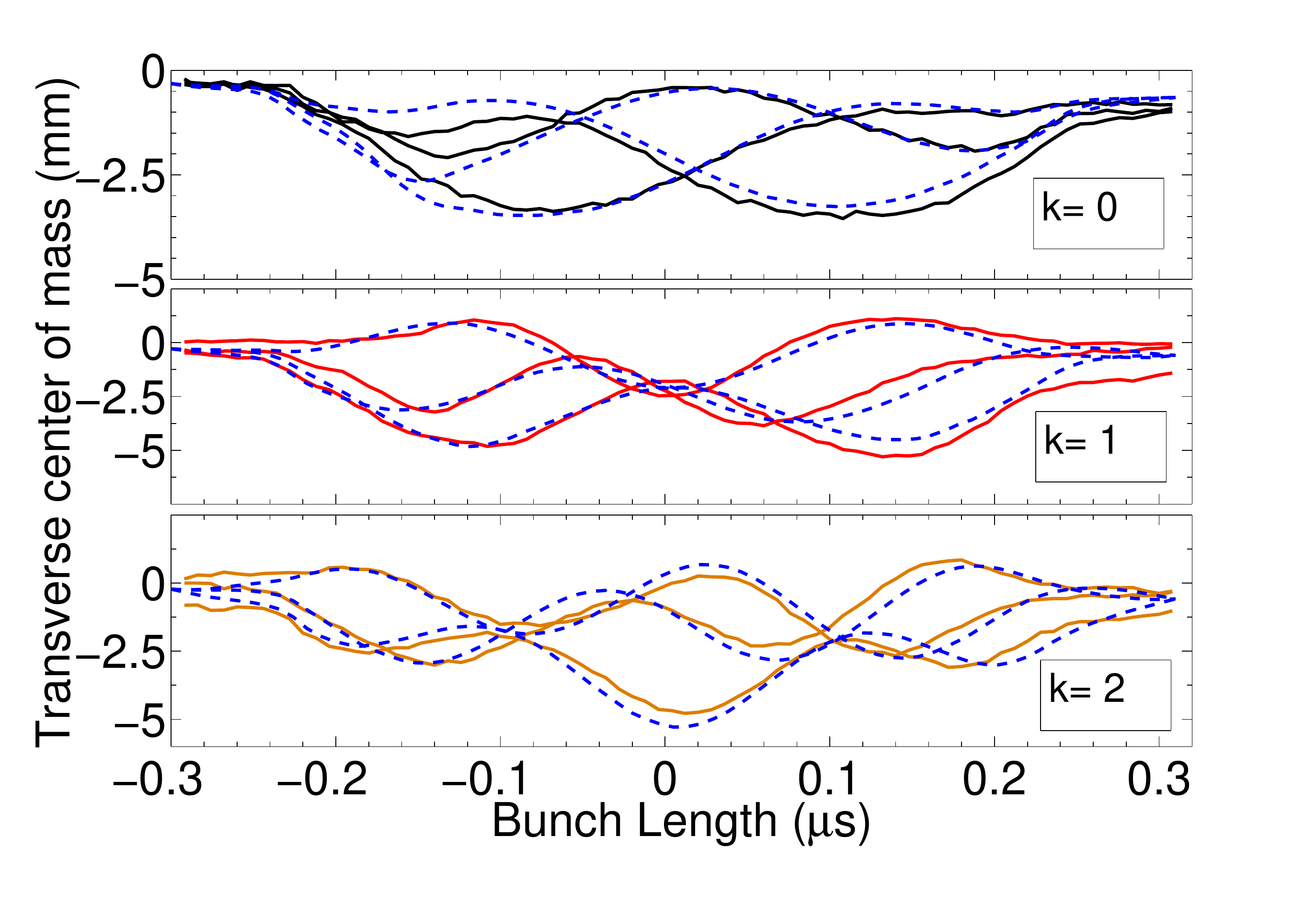}
   \caption{ Analytical curves from~\autoref{eq:eig_func} (dotted) are fitted to the local transverse offset for the $k=0$ and $k=1$ modes with the head-tail phase shift as the fit parameter. This method has been used for the precise determination of the chromaticity.}
   \label{fig:mode-fig1}
\end{figure}

As highlighted in the previous section, the frequency sweep allows to resolve the different head-tail modes both spectrally and temporally. This procedure can be used for the precise determination of the chromaticity by fitting 
the analytical expression for the head-tail eigenfunction~\autoref{eq:eig_func} to the measured bunch offset,
with the chromaticity ($\xi$) as the fit parameter as shown in~\autoref{fig:mode-fig1} for $k=0,1$ and $2$. The measured chromaticity is independent of the order of head-tail eigenfunction used to estimate it.

\begin{figure}[tbh]
\centering
   \includegraphics*[width=85mm]{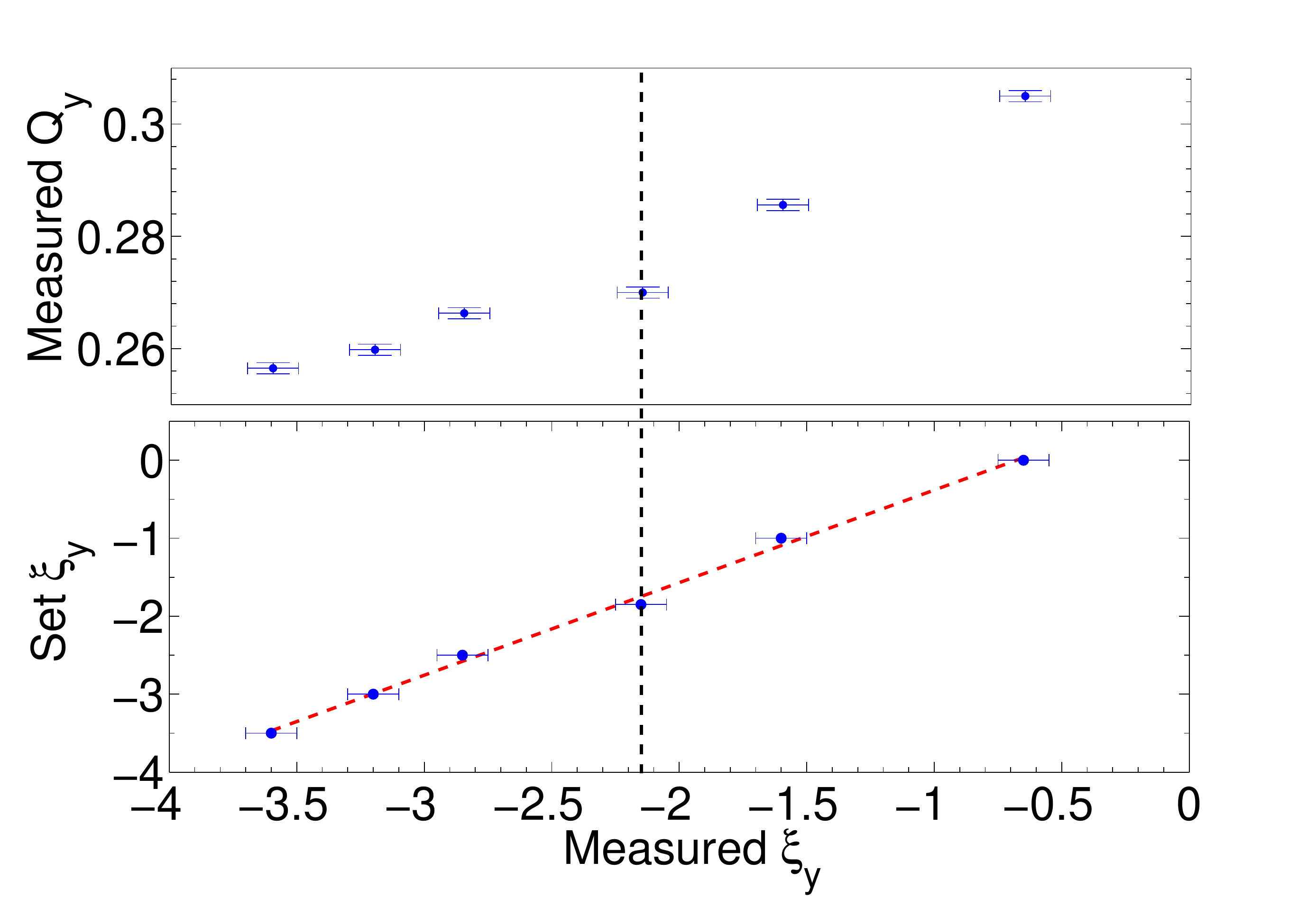}
   \caption{Chromaticity measurement using the method shown in~\autoref{fig:mode-fig} over the possible range of  operation at SIS-18. The black dashed line marks the measured natural vertical chromaticity of SIS-18. The error bars for the chromaticity measurement are determined mainly by the electronic noise in the amplifier chain of the TOPOS system.}
   \label{fig:chro_measure}
\end{figure}

The fitting method is shown in~\autoref{eq:chro_fit_proc}; the head-tail eigenfunction from~\autoref{eq:eig_func} is multiplied with the beam charge profile $\hat{S(t)}$ and corrected for the beam offset $\Delta{x}$ at the BPM where the signal is measured. 
\begin{eqnarray}\label{eq:chro_fit_proc}
F(\phi, \xi, A) = \Delta{x}\cdot \hat{S(t)}\cdot (1+A\cdot\bar{x}(\phi))\\ \nonumber
E(\xi, A) = (M(\phi) - F(\phi,\xi))^2
\end{eqnarray} 
The fit error $E(\xi, A)$ is reduced as a function of independent variables; chromaticity $\xi$ and head-tail mode amplitude $A$.  The fit error gives the goodness of the fit. It is used to determine the error bars on measured chromaticity.
This method has been utilized for the determination of chromaticity at SIS-18 as shown in~\autoref{fig:chro_measure}. The set and the measured chromaticity can be fitted by linear least squares to obtain the form $\xi_{s,y} = 1.187 \xi_{m,y}+0.804$ as shown by red dashed line in~\autoref{fig:chro_measure}.~\Autoref{fig:chro_measure} also shows a coherent tune shift due to change in sextupole strength which is used to adjust the chromaticity. This is due to uncorrected orbit distortions during these measurements.  These chromaticity measurements agree with the previous chromaticity measurements using conventional methods~\cite{Paret2009}. 

\begin{figure}[tbh]
\centering
   \includegraphics*[width=85mm]{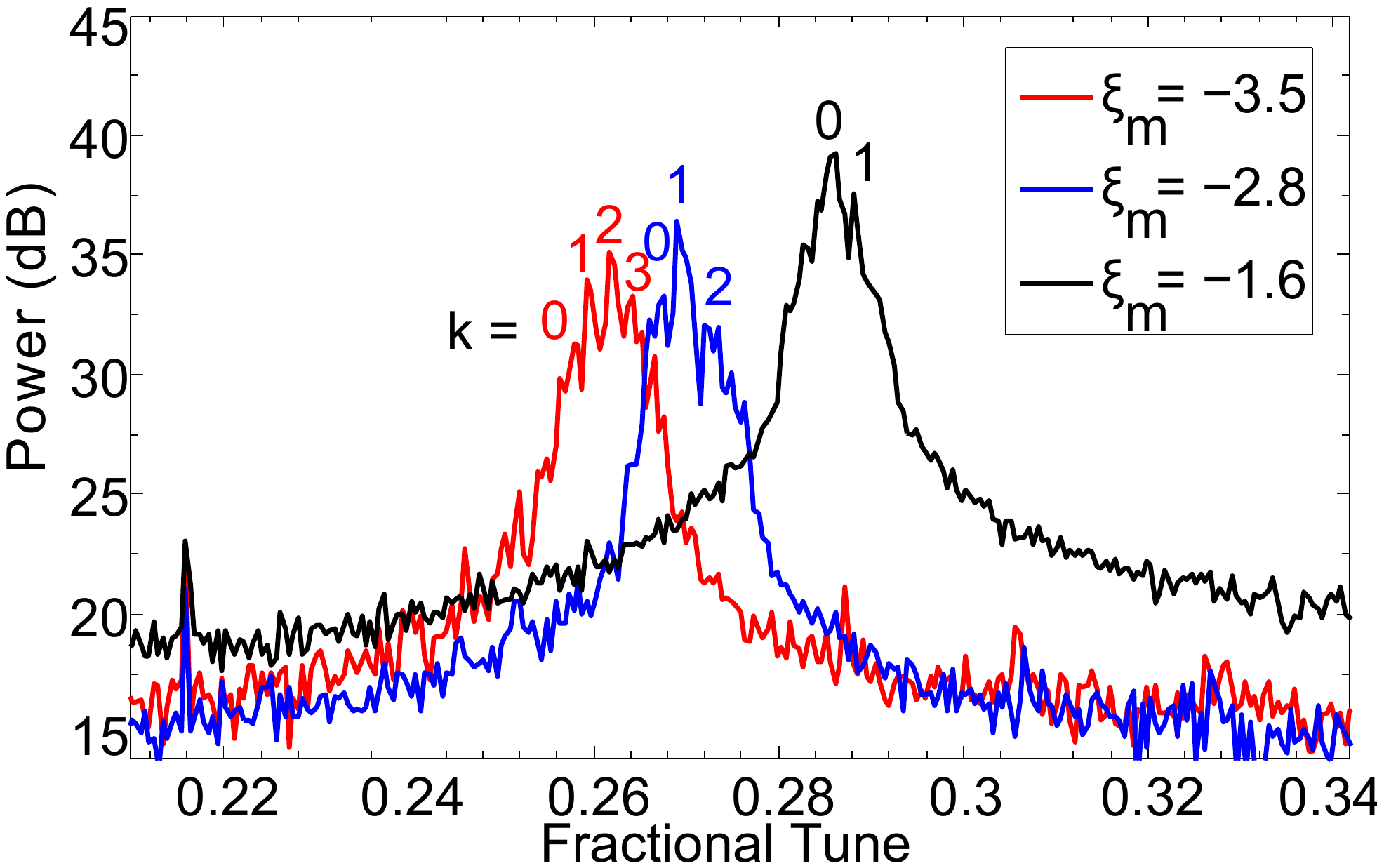}
   \caption{Relative signal amplitudes of various head-tail modes are plotted with respect to the measured chromaticity. The centroid of the spectrum moves toward higher order modes with increase in chromaticity.}
   \label{fig:chro-height}
\end{figure}

\begin{figure}[tbh]
\centering
   \includegraphics*[width=85mm]{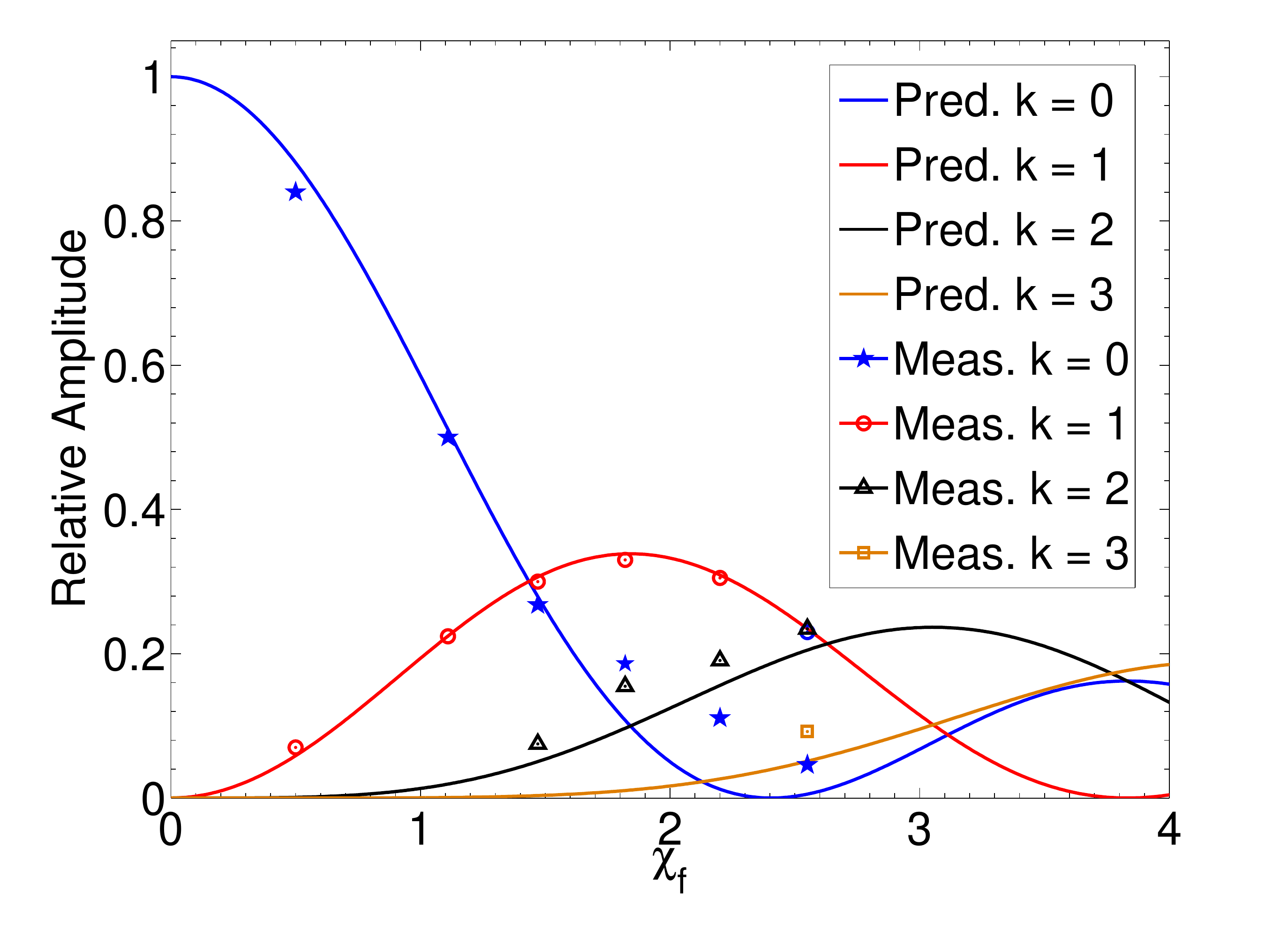}
   \caption{Relative signal amplitudes of various head-tail modes are plotted with respect to the coefficient of Bessel's function in~\autoref{eq:rel-amp} $(\chi_f = \chi/2)$. All the modes are relative to k=1 mode, which has a significant amplitude during the whole measurement range.}
   \label{fig:mode-amp}
\end{figure}

It is also possible to determine the relative response amplitude of each head-tail mode to the beam excitation both in time and in frequency domain with TOPOS.~\Autoref{fig:chro-height} shows the tune spectrum obtained with sweep excitation for different chromaticity values. The beam parameters are kept the same ($N^{7+},14\cdot 10^{9},q_{sc}\approx 10$). The spectral position and and relative amplitude of each head-tail mode peak are confirmed using the time domain information (see~\autoref{fig:mode-fig}). 
In \autoref{fig:mode-amp} the single particle response amplitudes for different $k$ (\autoref{eq:rel-amp}) are plotted as a function of the chromaticity. The measured relative amplitudes are indicated by the colored symbols. The comparison indicates that the simple single particle result (\autoref{eq:rel-amp}) describes quite well the dependence of the relative height of the peaks obtained from the TOPOS measurement.


\section{Application to tune measurements in SIS-18}\label{sec:discussion}

In this section we will discuss the application of our results to tune measurements in the SIS-18
and in the projected SIS-100, as part of the FAIR project at GSI \cite{FAIR2010}. As shown in 
the previous section the relative amplitudes of the synchrotron satellites in the tune spectra
are primarily a function of chromaticity and possibly the excitation mechanism. In order to determine the coherent tune with high precision the position of the $k=0$ mode has to 
be measured. Depending on the machine settings, if the relative height of the $k=0$ peak with respect to the other modes is small, then the $k=0$ mode may not be visible at all. To estimate the bare tune frequency in this case, the information of space charge parameter, coherent tune shift and chromaticity are all simultaneously required with good precision.

\begin{figure}[tbh]
\centering
   \includegraphics*[width=85mm]{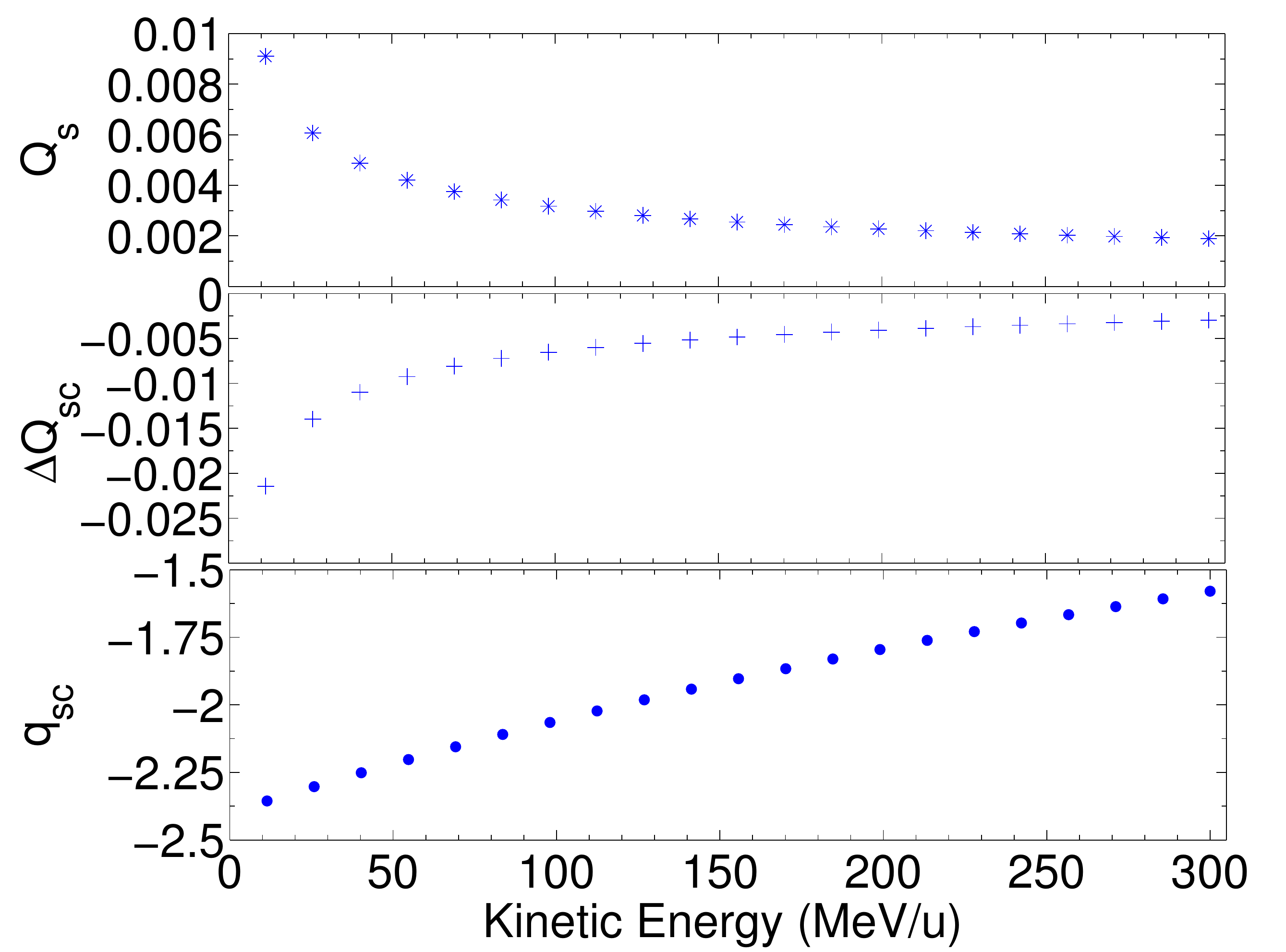}
   \caption{Change in synchrotron tune and space charge tune shift on acceleration from injection to extraction for $Ar^{18+}$ with $1\cdot10^{10}$ particles. The estimated space charge parameter reduces by $\approx 20\%$ in this typical case from injection at 11.4 MeV/u to 300 MeV/u at extraction.}
   \label{fig:energy_shift}
\end{figure}

\begin{figure}[tbh]
\centering
   \includegraphics*[width=85mm]{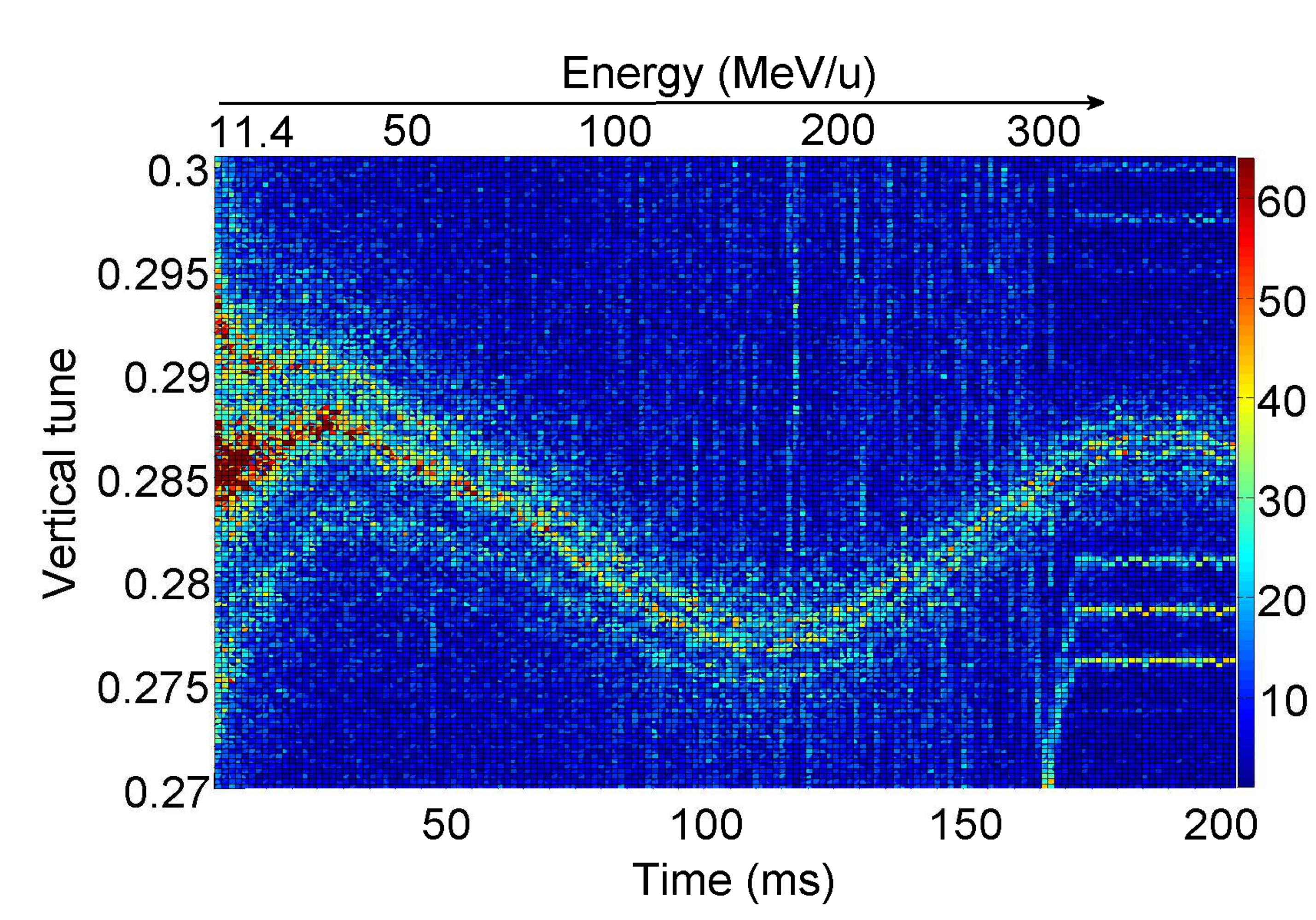}
   \caption{The movement of head-tail modes during acceleration. The head-tail modes are getting closer since the synchrotron tune reduces with acceleration, but the asymmetry of the $k=1$ and $k=-1$ modes around $k=0$ is maintained throughout the ramp which depends primarily on the space charge parameter $(q_{sc})$.}
   \label{fig:acc-tune-fig}
\end{figure}

Another important point is the tune measurement during acceleration.
The space charge parameter for $1\cdot 10^{10} Ar^{18+}$ stored ions in the SIS-18 from injection to extraction reduces only by $\approx 20\%$ as shown in~\autoref{fig:energy_shift}. The dynamic shift of head-tail modes during acceleration is shown in~\autoref{fig:acc-tune-fig} obtained from TOPOS system under same conditions. The asymmetry of $k=1,-1$ modes around $k=0$ mode can only be understood in view of the space charge effects predicted by~\autoref{eq:htmodes}. Thus, a correct estimate of this parameter plays an important role in understanding the tune spectra not only during dedicated experiments on injection plateau, but also during regular operations.

The measurement time required to resolve the various head-tail modes ($\Delta Q_k$) is a complex function of $Q_s$, $q_{sc}$, beam intensity and excitation power. To give some typical numbers for SIS-18; on a measurement time of 600 ms on the injection plateau, if one spectrum is obtained in $\approx 20$ ms ($\approx 4000$ turns), an improvement of factor $\approx 6$ in SNR by averaging 30 spectra. Following the calculations in~\autoref{subsec:inc_tune_shift}, $q_{sc} \lesssim 8$ can be resolved under typical injection operations. However, the constraints on measurement time are much higher during acceleration, where the tune/revolution frequency increases due to acceleration. This allows the measurement of a single spectrum typically only over $500-1000$ turns (depends on ramp rate as well). There are no averaging possibilities since the tune is moving during acceleration due to dynamic changes in machine settings as seen in~\autoref{fig:acc-tune-fig}. In addition, the synchrotron tune reduces with acceleration making it practically very difficult to resolve the fine structure of the head-tail modes for $q_{sc} \gtrsim 2$.

\section{Conclusion}

Two complimentary tune measurement systems, TOPOS and BBQ, installed in the SIS-18 are presented. 
Analytical as well as simulation models predict a characteristic modification of the tune spectra
due to space charge and image current effects in intense bunches. 
The position of the synchrotron satellites corresponds to the head-tail tune shifts and depends 
on the incoherent and the coherent tune shifts.  
The modification of the tune spectra for different bunch intensities has been observed in the SIS-18
at injection energy, using the TOPOS and BBQ systems.
From the measured spectra the coherent and incoherent tune shift for bunched beams in SIS-18 at injection energies were obtained experimentally using the analytic expression for the head-tail tune shifts.
Head-tail modes were individually excited and identified in time domain and correlated with the spectral information. A novel method for determination of chromaticity based on gated excitation of individual head tail mode is shown. The dependence of the relative amplitudes of various head-tail modes on chromaticity is also studied. The systems were compared against each other as well as with different kinds of excitation mechanisms and their respective powers. These measurements give a clear interpretation of tune spectra at all stages during acceleration under typical operating conditions. The understanding to tune spectra provides an important input to new developments related to planned transverse feedback systems for SIS-18 and SIS-100. The measurement systems also open new possibilities for detailed beam investigations as demonstrated in this contribution.

\section*{Acknowledgement}
We thank Marek Gasior of CERN BI Group for the help in installation of BBQ system and many helpful discussions on the subject. GSI operations team is also acknowledged for setting up the machine. We also thank Klaus-Peter Ningel from the GSI-RF group who helped setting up the amplitude ramp for the adiabatic bunching.

\section*{Appendix}

\subsection*{Calculation of bunching factor}
\begin{figure}[tbh]
\centering
   \includegraphics*[width=85mm]{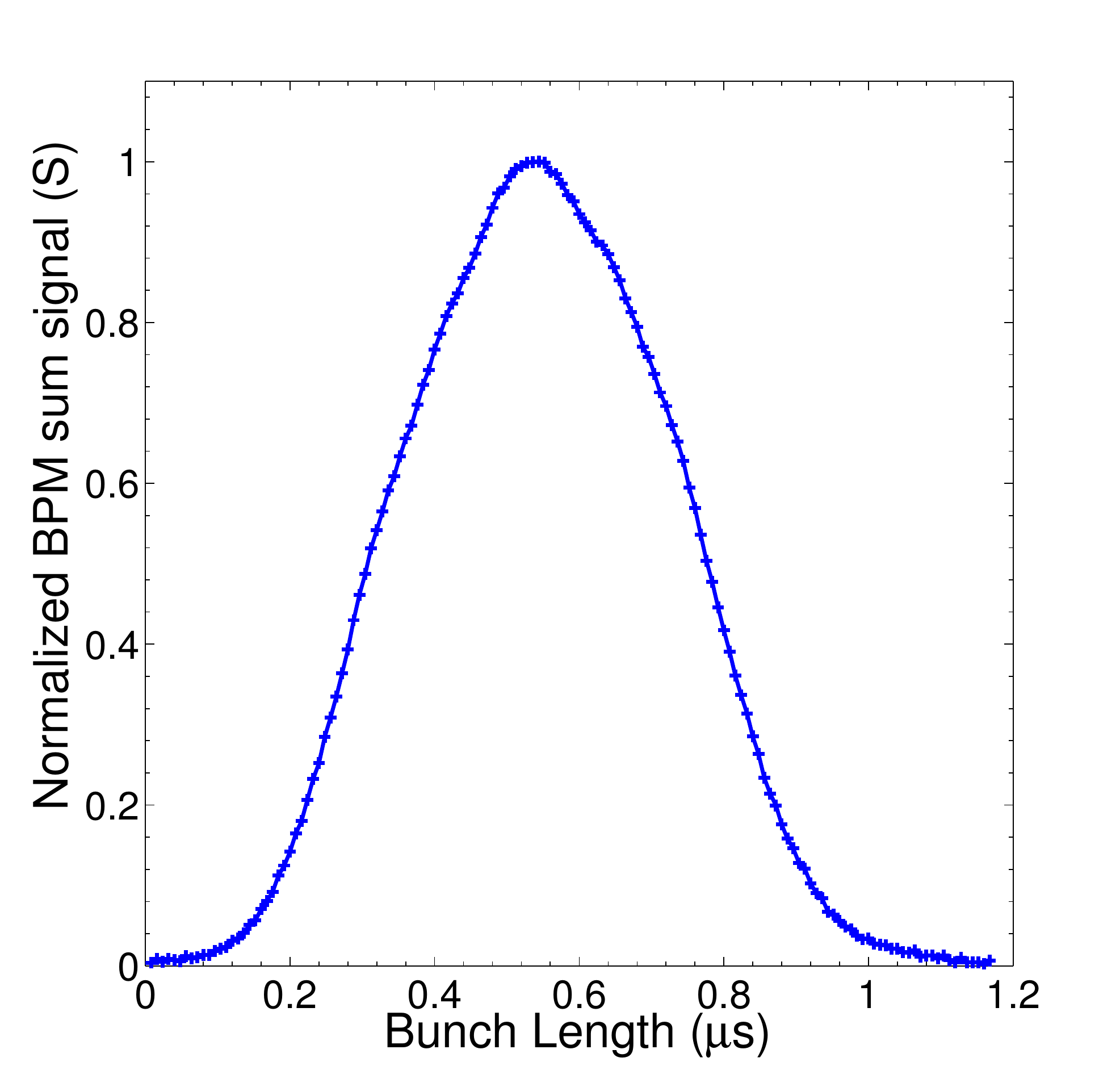}
   \caption{A typical normalized longitudinal beam profile from TOPOS system at injection. The length of the RF period is $\approx 1.2$  $\mu s$.}
   \label{fig:sum_bpm_signal}
\end{figure}

\Autoref{fig:sum_bpm_signal} shows a typical longitudinal profile of the bunch. If $S_j$ is the amplitude at each time instance $j=1,..,N$. The bunching factor is calculated by the~\autoref{eq:bunch_factor}.
\begin{equation*}\label{eq:bunch_factor}
\tag{A1}
B_f = \frac{\sum_{j=1}^N S_j}{max(S_j)\cdot N}
\end{equation*}
where N $\approx 147$ is the number of samples in one RF period (at injection). The TOPOS system samples the bunch at 125 MSa/s, thus the difference between adjacent samples is 8ns.

\subsection*{Measurement uncertainties}\label{subsec:meas_err}
The calculation of $\Delta Q_{sc}$ from~\autoref{eq:sctune} has a dependence on measured current, transverse beam profiles, longitudinal beam profiles and the twiss parameters. The measurement uncertainty on each of these measured parameters at GSI SIS-18 were commented in the detailed analysis in ~\cite{Franchetti2010}. Even though some parameters and the associated uncertainties are correlated, any correlations are neglected in present analysis. Uncertainties in each measured parameter are propagated to find the error bars on the calculated and measured incoherent tune shifts. 

Reproducing from Ref.~\cite{Franchetti2010}, the relative random uncertainty (std. deviation) in beam profile width($\sigma_{x}$) measurements is given by~\autoref{eq:ran_ipm}.

\begin{equation*}\label{eq:ran_ipm}
\tag{A2}
\frac{\delta \sigma_x}{\sigma_x} = \frac{0.043}{B} + 0.33\delta N_{i}
\end{equation*}

where $\Delta x = 2.1 $mm is the wire spacing and $\delta N_{i} =\frac{1}{2^8}$ is the ADC resolution of the IPM and B is defined as $\sigma_{x}/\Delta x$.
If the error bars are derived from $j$ measurements, the measured profile is given by
\begin{equation*}\label{eq:many_meas}
\tag{A3}
\sigma_{av,x} = \langle \sigma_x\rangle_j, \delta\sigma_{av,x} =\sqrt{{\langle \sigma_x^2\rangle}_j -\langle \sigma_x\rangle^2_j+\langle{\delta}{\sigma_x\rangle}^2}
\end{equation*}
For each tune measurement at the given intensity and excitation power, 5-8 transverse beam profiles were measured, and the relative error is obtained $\approx 5\%$ using~\autoref{eq:ran_ipm} and~\autoref{eq:many_meas}. The relative systematic error (bias) in transverse beam width measurements is $< 1\%$~\cite{Franchetti2010} and ignored in this analysis.

The uncertainty in the injected current is dominated by fluctuations in the source and the relative uncertainty is estimated to be $\approx 5\%$ based on 5-8 measurements at the same intensity settings for each measurement point. Bunch length and bunching factor vary by $\approx 2-3\%$ due to long term beam losses only under high intensity beam conditions. The maximum relative bias in the lattice parameter $\beta$ is assumed to be $\approx 5\%$ at the IPM location. Taking all the relative errors, uncertainty propagation using familiar~\autoref{eq:prop_err} gives relative error for estimated incoherent tune shifts $\approx 12\%$.
\begin{align*}\label{eq:prop_err}
\tag{A4}
&\frac{\delta \varepsilon_{av,x}}{\varepsilon_{av,x}} = \sqrt{4{(\frac{\delta \sigma_{av,x}}{\sigma_{av,x}})}^2+{(\frac{\delta \beta_{x}}{\beta_x})}^2} \nonumber \\
&\frac{\delta \Delta Q_{sc}}{\Delta Q_{sc}} = \sqrt{{(\frac{\delta \varepsilon_{av,x}}{\varepsilon_{av,x}})}^2+{(\frac{\delta I_p}{I_p})}^2}
\end{align*}

Tune measurements done by averaging over long intervals contribute to the width of modes due to long term beam losses. Beam losses lead to change in coherent tune especially in the vertical plane where the image current effects are larger. This has been highlighted at appropriate sections in the text.

%\bibliography{draft_PRST_mode_shift}% Produces the bibliography via BibTeX.

\end{document}